\documentclass[letterpaper,twocolumn,10pt]{article}
\usepackage{usenix-2020-09}

\usepackage{tikz}
\usepackage{amsmath}
\usepackage{subcaption}
\usepackage{graphicx}
\usepackage{graphics} 
\usepackage{filecontents}
\usepackage{amssymb}
\usepackage{mdframed}
\usepackage{multirow}
\usepackage{booktabs}
\usepackage{threeparttable} 
\makeatletter

\newcommand{\Rmnum}[1]{\expandafter\@slowromancap\romannumeral #1@}
\makeatother

\usepackage{float}
\usepackage{bbm}
\usepackage{cite}
\usepackage{colortbl}
\usepackage{makecell} 
\usepackage{xcolor}
\definecolor{mygray}{gray}{.9}
\usepackage[toc,page]{appendix}
\usepackage{amsthm}
\newtheorem{theorem}{Theorem}[section]
\newtheorem{lemma}[theorem]{Lemma}

 \usepackage[available]{usenixbadges}

\begin{document}
\title{SoK: On Gradient Leakage in Federated Learning}
%Get back to basics

\author{{\rm Jiacheng Du$^{\dagger, \wr,\sharp}$ \quad Jiahui Hu$^{\dagger, \wr,\sharp}$ \quad Zhibo Wang$^{\dagger, \wr, \ast}$}\\ {\rm Peng Sun$^{\natural}$ \quad Neil Zhenqiang Gong$^{\ddagger}$ \quad Kui Ren$^{\dagger, \wr}$ \quad Chun Chen$^{\dagger, \wr}$} \\
{\small$^{\dagger}$The State Key Laboratory of Blockchain and Data Security, Zhejiang University, P. R. China}\\
{\small$^{\wr}$Hangzhou High-Tech Zone (Binjiang) Institute of Blockchain and Data Security, P. R. China} \\
{\small$^{\natural}$College of Computer Science and Electronic Engineering, Hunan University, P. R. China}\\
{\small$^{\ddagger}$Department of Electrical and Computer Engineering, Duke University, USA}\\
% {\small$^{\wr}$ZJU-Hangzhou Global Scientific and Technological Innovation Center} \\
% {\small$^{\sharp}$College of Computer Science and Electronic Engineering, Hunan University, P. R. China} \\ 
% {\small$^{\natural}$School of Cyber Science and Engineering, Wuhan University, P. R. China}\\
{\tt\small \{jcdu, jiahuihu, zhibowang, kuiren, chenc\}@zju.edu.cn, psun@hnu.edu.cn, neil.gong@duke.edu}
}
\maketitle
\newcommand\blfootnote[1]{%
\begingroup
\renewcommand\thefootnote{}\footnote{#1}%
\addtocounter{footnote}{-1}%
\endgroup
}
\blfootnote{$^\ast$Zhibo Wang is the corresponding author.}
\blfootnote{$^\sharp$Jiacheng Du and Jiahui Hu contribute equally to this work.}

\maketitle

\begin{abstract}
Federated learning (FL) facilitates collaborative model training among multiple clients without raw data exposure. However, recent studies have shown that clients' private training data can be reconstructed from shared gradients in FL, a vulnerability known as gradient inversion attacks (GIAs). While GIAs have demonstrated effectiveness under \emph{ideal settings and auxiliary assumptions}, their actual efficacy against \emph{practical FL systems} remains under-explored. To address this gap, we conduct a comprehensive study on GIAs in this work. We start with a survey of GIAs that establishes a timeline to trace their evolution and develops a systematization to uncover their inherent threats. By rethinking GIA in practical FL systems, three fundamental aspects influencing GIA's effectiveness are identified: \textit{training setup}, \textit{model}, and \textit{post-processing}. Guided by these aspects, we perform extensive theoretical and empirical evaluations of SOTA GIAs across diverse settings. Our findings highlight that GIA is notably \textit{constrained}, \textit{fragile}, and \textit{easily defensible}. Specifically, GIAs exhibit inherent limitations against practical local training settings. Additionally, their effectiveness is highly sensitive to the trained model, and even simple post-processing techniques applied to gradients can serve as effective defenses. Our work provides crucial insights into the limited threats of GIAs in practical FL systems. By rectifying prior misconceptions, we hope to inspire more accurate and realistic investigations on this topic.

\end{abstract}


% \fi
%\IEEEpeerreviewmaketitle

\section{Introduction}
\label{sec:intro}
%-------------------------------------------------------------------------------
Federated learning (FL)\cite{mcmahan2017communication} has recently become a widely adopted privacy-preserving distributed machine learning paradigm. In FL, multiple clients collaborate to train a global model orchestrated by a central server for multiple rounds. In each round, clients update the global model locally using private training data and then transmit gradients to the server for aggregation and global update, thereby alleviating privacy concerns from raw data exposure. Therefore, FL has attracted considerable academic interest and empowered various real-world applications, including mobile services such as Google Keyboard\cite{hard2018federated}, healthcare\cite{li2021federated}, and finance\cite{long2020federated}.\par

\begin{figure*}[t]
    \centering
    \vspace{-2mm}
    \includegraphics[width=1\textwidth]{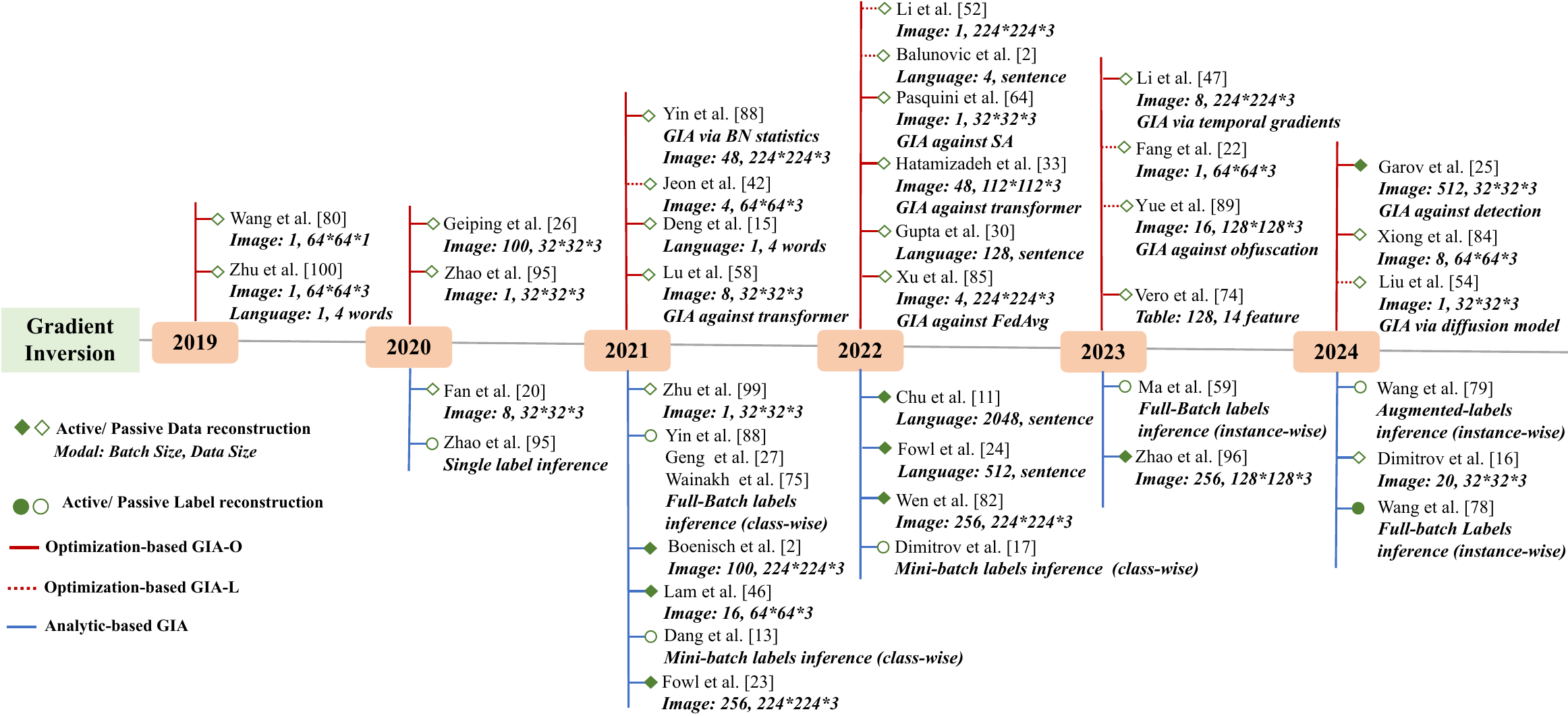}

    \caption{Evolution of Gradient Inversion Attack.}
    \label{fig:milestones}
    \vspace{-3mm}
\end{figure*}
\textbf{However, recent works claim that clients' data privacy can be compromised by their gradients sharing in FL.}\cite{wang2019beyond,zhu2019deep} Notably, a curious central server can reconstruct their private data by employing gradient inversion attacks (GIAs)\cite{wang2019beyond}. Fig.~\ref{fig:milestones} depicts the development and milestones of the GIA, presenting two forms:\par
\textbf{Optimization-based GIA:} GIAs typically assume an \textit{honest-but-curious} server and employ optimization-based methods to \textbf{passively} reconstruct the victim client's training data \cite{wang2019beyond, zhu2019deep}. In this approach,  the adversary (server) randomly initializes data and labels, computes gradients based on the same model as the victim client, and iteratively updates these initializations to mirror the ground truth by minimizing the distance between the computed gradient and the client's shared gradient\cite{zhu2019deep}. Further, it can be categorized into two primary subforms based on the optimization space: GIA with observable space optimization (GIA-O)\cite{geiping2020inverting,yin2021see,hatamizadeh2022gradvit,gupta2022recovering,lu2022april,liu2022breaking} and GIA with latent space optimization (GIA-L)\cite{jeon2021gradient,li2022auditing,balunovic2022lamp}, as defined in Sec.~\ref{sec:sys}. Recent advancements have enabled GIAs to reconstruct larger-batch or higher-dimension data\cite{yin2021see,jeon2021gradient}, invert more complex model architectures\cite{hatamizadeh2022gradvit,balunovic2022lamp,chu2022panning}, and adapt to various tasks\cite{li2022auditing,vero2022data} and FL protocols\cite{pasquini2022eluding,xu2022agic,wang2024breaking}.\par
\textbf{Analytic-based GIA:} Analytic-based methods aim to directly reconstruct training data and labels by formulating and solving equations between gradients and inputs \cite{fan2020rethinking, zhu2020r}. Initially, these methods were primarily used to infer labels directly from gradients\cite{zhao2020idlg,yin2021see}, but were limited to inverting low-dimension data on shallow models\cite{fan2020rethinking, zhu2020r}. To reconstruct higher-dimension inputs, recent efforts have assumed a \textit{malicious} server capable of \textbf{actively} crafting~\cite{fowl2021robbing,fowl2022decepticons,zhao2023resource} or modifying~\cite{boenisch2023curious,wen2022fishing} model parameters. Consequently, when a client trains the \textit{malicious} model, the training data leave an ``imprint'' in the shared gradient, enabling the server to retrieve them by solving equations.\par

Despite the rapid growth and impressive performance, \textbf{there remains skepticism about GIA's real capability and threat to practical FL systems, as often claimed}. \textit{On one hand, existing works tend to obsess over employing auxiliary assumptions to achieve heightened performance.} Reviewing the milestones, Yin et al. \cite{yin2021see} assumed the adversary possesses additional access to BN statistics as side information, while Lam et al. \cite{lam2021gradient} relaxed server assumptions to a malicious extent, allowing the adversary to impractically tamper with the model. \textit{On the other hand, these works often evaluate GIAs in settings far from practicality}. For instance, literature often assumes a specific client who aggregates all private data into a single batch and updates the model for one step, enabling the adversary to obtain the raw gradients. However, in practical scenarios, the shared gradients are model updates after mini-batch Stochastic Gradient Descent (SGD) and multiple-epoch training \cite{mcmahan2017communication}. Moreover, the target models are often specially initialized \cite{zhu2019deep} or designed \cite{geiping2020inverting, yin2021see} solely to achieve superior reconstruction quality from gradients.\par

In this work, we conduct a comprehensive study on GIA to better understand its development, properties, and real threats to practical FL systems. Specifically, we make the following key contributions:\par
\textbf{\Rmnum{1}. A survey on GIA.} We start by establishing a summary on \textit{the development of the GIA}. We thoroughly review the related works on the GIA, highlighting the milestones and breakthroughs in performance, as shown in Fig.~\ref{fig:milestones}. Moreover, we conduct a \textit{systematization of the GIA} along three dimensions (Sec.~\ref{sec:sys}): \textit{threat model}, \textit{attack}, and \textit{defense}, as detailed in Tab.~\ref{tab:sys} in Appx~\ref{appx:sup}. Notably, we characterize the threat model of the GIA, categorizing the assumptions based on their practical accessibility to potential adversaries.% Moreover, through our survey, our subsequent study focuses on \textit{optimization-based} GIAs with \textbf{honest-but-curious} server on \textbf{image reconstruction} task, because they are more practical and applicable in FL system compared with \textbf{analytic-based GIAs}. 

\textbf{\Rmnum{2}. Extensive investigations on GIA in practical FL.} We identify \textit{three fundamental aspects} that influence GIA's effectiveness in FL: \textit{training setup}, \textit{model}, and \textit{post-processing} (Sec.~\ref{rethinking}). Subsequently, we conduct comprehensive theoretical analyses and empirical evaluations on GIAs from the aspects (Sec.~\ref{sec:training}, Sec.~\ref{sec:model}, and Sec.~\ref{sec:post-processing}) across diverse settings. Our investigations bridge the gap between literature and practice, revealing the real threats posed by GIAs in FL systems.\par

\textbf{\Rmnum{3}. Analyses and insights for GIA in practical FL.} We provide an in-depth analysis of GIA's properties and the associated real threats, offering a summary of key insights. Our findings indicate that, despite its purported effectiveness, GIA is notably \textit{constrained, fragile, and easily defensible}, as supported by the following observations:\par
\textit{\textbf{(a) GIA presents inevitable bottlenecks against practical training setups.}} We investigate how the client's local training impacts GIAs from training configuration and training data. Specifically, we theoretically prove that \textit{as the number of local updates increases, reconstruction becomes much more difficult} (Sec.~\ref{sec:training-1}). Moreover, we evaluate the capabilities of SOTA GIAs across a wide range of data dimensions. With theoretical proofs, we indicate the \textit{GIA's bottleneck in data reconstruction as data dimension growth} (Sec.~\ref{sec:training-2-1}). Besides, \textit{we identify that data content significantly affects GIAs with two canary cases} (Sec.~\ref{sec:training-2-2}). Specifically, GIA fails to reconstruct semantic details containing crucial private information. For generative adversaries, out-of-distribution (OOD) data constrains GIA's performance.\par

\textit{\textbf{(b) GIA is extremely sensitive to the model, including training stage and architecture.}} We propose a novel \textbf{I}nput-\textbf{G}radient \textbf{S}moothness \textbf{A}nalysis (\textbf{IGSA}) method to quantify and explain the model's vulnerability to the GIA during the FL training process (Sec.~\ref{sec:model-1}). Surprisingly, our findings indicate that \textit{the GIA exclusively works in the early training stages}. Furthermore, we undertake \textit{a deep investigation into the strong correlation between model architecture and the GIA}. Our analyses highlight that commonly employed structures (e.g., skip connections\cite{he2016deep}), and even seemingly insignificant micro designs (e.g., ReLU), significantly impact the model's resilience against GIAs (Sec.~\ref{sec:model-2}).\par
%\textbf{(2)} \textbf{GIA is extremely sensitive to the model, on both training stage and architecture.} We propose a novel \textbf{I}nput-\textbf{G}radient \textbf{S}moothness \textbf{A}nalysis (\textbf{IGSA}) method to quantify and explain model's vulnerability during FL training process (Sec.\ref{sec:model-1}). And we are surprised to find that \textit{GIA works only in the early training stages}. Besides, we are \textit{the first to investigate the strong association between model architecture and GIA}. Our theoretical and empirical analyses indicate that the commonly used structures (e.g. skip connection\cite{he2016deep}, and Net-In-Net (NIN)\cite{{szegedy2015going}}), even the insignificant micro-designs(e.g., Relu, kernel function size, etc.), have profound effects on model's resistance to GIA (Sec.\ref{sec:model-2}).\par
\textit{\textbf{(c) Even trivial post-processing measures applied to gradients in practical FL systems can effectively defend against GIAs while maintaining model accuracy.}} We evaluate four post-processing techniques, considering the privacy-utility trade-off within a practical FL setting. We show that \textit{even when confronted with SOTA GIAs, clients can readily defend against them by employing post-processing strategies (e.g., quantization~\cite{alistarh2017qsgd}) to obscure the shared gradients} (Sec.~\ref{sec:post-processing}).

\section{Systematization of Gradient Inversion}
\label{sec:sys}
%In this section, we conduct a systematization of gradient inversion in terms of four dimensions: \textbf{\textit{threat model}}, \textbf{\textit{attack}} and \textbf{\textit{defense}}. Due to the space limitation, we summarize all the related works to date based on above dimensions, as Tab.~~ \ref{tab:sys} in Appx.\ref{appx:sup} shows.
\subsection{System Model}
%\neil{Split into shorter pararaphs} 
We consider an FL system consisting of a server and $M$ clients, each with a private training dataset containing $N$ pairs of data ($\mathbf{x}$) and labels ($\mathbf{y}$). The $M$ clients collaboratively train a global model $W^{g}$ over $T$ rounds under the coordination of the server. In each round $t$, the server selects $K$ clients and sends the current global model $W^{g}_{t}$ to them for local training. Each client $k$ performs $E$ epochs of local training utilizing mini-batch SGD with a batch size of $B$. Consequently, each client performs $U = EN/B$ local updates. The different configurations of $E$ and $B$ give rise to two FL protocols:\par
(1) \textbf{FedSGD} (Federated Stochastic Gradient Descent) \cite{konevcny2015federated}: Each client $k$ aggregates all $N$ local training data samples into a batch ($B = N$) and executes a single epoch of local training ($E = 1$), updating $W^{g}_{t}$ for one time. The computed gradient $\nabla_{k} W^{g}_{t}$ is uploaded to the \textit{server}. The \textit{server} aggregates the collected gradients and updates the global model as follows: $W^{g}_{t+1} \leftarrow W^{g}_{t} + \eta\sum_{k=1}^K\nabla_{k} W^{g}_{t}$.\par
(2) \textbf{FedAvg} (Federated Averaging) \cite{mcmahan2017communication}: Client $k$ conducts $E>1$ epochs of mini-batch SGD with $B\le N$ for local training and shares an updated local model $W^{k}_{t+1}$. The server receives the $K$ local models and computes their average to obtain the updated global model: $W^{g}_{t+1} \leftarrow \frac{1}{K}\sum_{k=1}^KW^{k}_{t+1}$. 

For the server, the primary distinction between the two protocols lies in the type of parameters shared. In FedSGD, the server directly receives the \textbf{gradient}, which captures the precise parameter changes of $W^{g}_{t}$ after a one-step gradient descent. In contrast, clients share the updated models in FedAvg. Consequently, the server could only obtain the parameter \textbf{updates} by $W^{k}_{t+1}-W^{g}_{t}$, which are the cumulative parameter changes after multiple steps of mini-batch gradient descent.

\subsection{Threat Model}
Existing studies primarily focus on scenarios where the server, receiving gradients from clients, acts as the adversary conducting GIAs. Therefore, we conduct a summary on the threat model of existing GIAs based on the following three aspects: \par
%We mainly focus on the adversary comes from the server that receives the shared gradients. Therefore, we sort out the threat model in terms of the \textit{goal}, \textit{capability} (with \textit{server's trustworthiness}) and \textit{assumption} of the adversary: \par %In particular, our work is the first to summarize and rank the assumptions according to their accessibility. We hope that a detailed discussion of existing threat models will inspire future works, exploring the privacy threats posed by gradients in a reasonable context.\par
\textit{(1) Goal.} The goal of the adversary is to reconstruct the client's \textbf{data} and \textbf{labels} from the shared gradients. Initial investigations  (e.g., \cite{zhu2019deep}) attempt to reconstruct the data and labels simultaneously. However, follow-up works\cite{zhao2020idlg,yin2021see} reveal that labels could be inferred directly from the gradients without explicitly solving them. As a result, subsequent GIAs focus on reconstructing either labels or data separately. Most GIAs concentrate on data reconstruction, aiming to retrieve more accurate private data, assuming that labels are either already known or can be reliably inferred beforehand. Meanwhile, investigations such as \cite{dang2021revealing,dimitrov2022data,ma2022instance} explore label inference, which has two implications: First, foreknowledge of labels aids subsequent data reconstruction. Second, labels themselves contain sensitive information such as a user's purchase history\cite{li2021label}. So far, GIAs have been capable of inferring the presence of certain classes (class-wise)\cite{yin2021see} and further determining the number of instances within each class (instance-wise)\cite{ma2022instance} from a full-batch\cite{yin2021see,geng2021towards,wainakh2021user,ma2022instance,wang2024towards} or multiple mini-batches of data\cite{dang2021revealing,dimitrov2022data}. \par

\textit{(2) Capacity \& Server's Trustworthiness.} Most existing GIAs presume an \textbf{honest-but-curious} server\cite{wang2019beyond,zhu2019deep,zhao2020idlg,geiping2020inverting,yin2021see,fan2020rethinking,zhu2020r,geng2021towards,wainakh2021user,deng2021tag,jeon2021gradient,dang2021revealing,li2022auditing,hatamizadeh2022gradvit,lu2022april,balunovic2022lamp,gupta2022recovering,dimitrov2022data,xu2022agic,ma2022instance,li2023temporal}, which implies that the server merely analyzes shared gradients \textbf{passively} without interrupting the training process. Consequently, the victim client remains unaware, ensuring the stealthiness of GIAs. Furthermore, some studies consider that a \textbf{malicious} server not only analyzes the gradient but could also \textbf{actively} interfere with the learning process through malicious behaviors, thereby extracting more information about the input by gradients. Specifically, they consider that the server can craft\cite{fowl2021robbing,chu2022panning,pasquini2022eluding} or modify\cite{fowl2022decepticons,wen2022fishing,zhao2023resource,boenisch2023curious} the model parameters, enabling the adversary to achieve better reconstruction results. Besides, unlike the \textbf{honest-but-curious} server, such \textbf{malicious} behaviors could be easily detectable by clients.\par

\textit{(3) Assumption.} Assumptions specify the adversary's knowledge and allowed behaviors. Reflecting on the evolution of the GIA, diverse assumptions offer additional advantages to the adversary and even serve as the key to the asserted impressive performances\cite{huang2021evaluating}. Herein, we categorize and rank existing assumptions based on their accessibility to the adversary within practical FL systems and access GIAs in Tab.~\ref{tab:sys}.\par
%\textit{(3) Assumption.} Looking back at the development of GIA, literature has developed various assumptions, which specify adversary's knowledge and behaviors. Assumptions provide additional aids to the adversary and even become key to the claimed performances of some GIAs. Here, we rank existing assumptions according to the accessibility of the adversary in practical FL systems and mark them for each work in Tab.~\ref{tab:sys}:\par
\textbf{[Level $0$]: Basic information} refers to the necessities for gradient inversion, including gradient, model, data dimension, and number of data samples $N$ at victim clients, which are readily accessible to the server in practical FL systems.\par
\textbf{[Level $1$]: Priors} refer to side information, including established patterns or observations that are readily accessible. For instance, Geiping et al. in\cite{geiping2020inverting} utilized total variation \cite{rudin1992nonlinear} as a prior, an established pattern of smoothness among neighboring pixels, effectively regulating the reconstruction process. Additionally, certain priors stem from observations on gradients, for example, Lu et al. in\cite{lu2022april} discovered that the cosine similarity of gradients in the positional embedding layer is substantial for two similar images. Consequently, they designed a regularization term to invert vision transformers\cite{dosovitskiy2020image}.\par
% Yin et al.\cite{yin2021see} assume that the reconstructed data should satisfy a group consistent registration prior in order to reduce the randomness of the initializations.
\textbf{[Level $2$]: Data distribution} refers to the statistical characteristics of the client's private dataset. Knowing the distribution enables the adversary to pre-train a generator, thereby improving the data reconstruction performance\cite{wang2019beyond,jeon2021gradient,li2022auditing,balunovic2022lamp,fang2023gifd,xiong2024gi,yue2023gradient} (further elaborated in Eq.~\eqref{eq2}). In FL, clients are not mandated to disclose their data distribution to the server. Nevertheless, given the server's approximate knowledge of the task, it may occasionally estimate the distribution using open-source datasets. For instance, if the server knows that the client possesses facial data, it may utilize datasets like FFHQ\cite{karras2019style} for generator pre-training.\par
\textbf{[Level $3$]: Client-side training details} refer to settings such as local learning rates, epochs, mini-batch sizes, optimizer, etc. Xu et al. in\cite{xu2022agic} introduced a GIA capable of quickly approximating client-side multi-step updates in FedAvg, necessitating access to these training details. However, such information is typically unavailable to the server.\par
\textbf{[Level $4$]: BN statistics} are the mean and variance of batched data acquired at BN layers that may be uploaded along with gradients in FL. They were initially utilized in the GIA by Yin et al.\cite{yin2021see} and subsequently in several works\cite{hatamizadeh2022gradvit,huang2021evaluating,xu2022agic,li2023temporal}. However, in practice, the BN layer can be easily substituted\cite{wu2018group}, and clients typically are not required to upload their BN statistics\cite{li2021fedbn}.\par
\textbf{[Level $5$]: Malicious behavior} involves an adversary's active manipulation of protocols\cite{lam2021gradient}, models and other components in FL to enhance the 
reconstruction performance. Existing studies concentrate on crafting\cite{fowl2021robbing,chu2022panning,pasquini2022eluding} or modifying\cite{fowl2022decepticons,wen2022fishing,zhao2023resource,boenisch2023curious,zhao2023loki} model parameters. Notably, recent works\cite{fowl2021robbing,boenisch2023curious} presume the existence of a \textbf{large fully connected layer} at the \textbf{front} of the model, yet such anomalous designs can be easily detected\cite{garov2023hiding}. Consequently, these behaviors lack stealthiness and practical applicability.\par
In summary, \textbf{Level 0} and \textbf{1} are assumptions easily accessible to the adversary in practical FL systems. On the other hand, \textbf{Level 2}, \textbf{3}, and \textbf{4} are deemed strong assumptions as practical clients are not required to furnish this information to the server, although the adversary might approximate or acquire it under certain circumstances. Additionally, we contend that \textbf{malicious behavior (Level 5)} is over-assumed in practical FL systems due to its inherent lack of stealthiness. 

\subsection{Attack}
GIAs employ two main attack strategies and involve several modalities based on the different types of FL tasks.

\textit{(1) Strategy.} The attack strategies employed by GIAs can be categorized into two forms: \textbf{optimization-based} and \textbf{analytic-based}, as illustrated in Fig.~\ref{fig:milestones}.\par
The process of \textbf{optimization-based} GIA involves iteration from a random initialization towards an approximation of the ground truth data, guided by a loss function of gradient similarity. It can be further categorized into two primary sub-forms based on the optimization space:\par
\textbf{1) GIA-O:} At training round $t$, the victim client holding $N$ pairs of data $\mathbf{x}$ and labels $\mathbf{y}$ shares its gradient $\nabla W$ to the server after local training ($B\leq N$, $U\geq 1$). The adversary obtains $\nabla W$ and generates $N$ pairs of randomly initialized data $x^{\prime}$ and labels $y^{\prime}$ with the same dimension of ground truth. Following the loss of gradient similarity, $N$ pairs of initialization are updated until they approximate the ground truth ($\mathbf{x}$, $\mathbf{y}$) pairs: 
\newtheorem{definition}{Definition}
\begin{definition}[GIA with Observable Space Optimization]
\begin{equation}\label{eq1}
\begin{aligned}
& \{\mathbf{x}_{n}^{\prime*}, \mathbf{y}_{n}^{\prime*}\}_{n=1}^N = \\
& \operatorname*{\text{argmin}}_{\{\mathbf{x}_{n}^{\prime}, \mathbf{y}_{n}^{\prime}\}_{n=1}^N} \mathrm{Dist}&\left( \frac{1}{N}\sum_{n=1}^N \frac{\partial \ell(\mathbf{x}_{n}^{\prime},\mathbf{y}_{n}^{\prime})}{\partial W}  - \nabla W\right) + \alpha \mathcal{R},
\end{aligned}
\end{equation}
where $\mathrm{Dist}\left(\cdot\right)$ is a distance metric between two vectors (such as Euclidean distance\cite{zhu2019deep,wang2019beyond} and cosine similarity\cite{geiping2020inverting}), $\mathcal{R}$ represents regularization terms, e.g., total variation \cite{rudin1992nonlinear} and BN statistics\cite{yin2021see}, and $\alpha$ is the weighting factor.\par
% \begin{equation}\label{eq1}
% \begin{aligned}
%  \{x_{n}^{\prime*}, y_{n}^{\prime*}\}_{n=1}^N = \operatorname*{\text{arg min}}_{\{x_{n}^{\prime}, y_{n}^{\prime}\}_{n=1}^N} \mathrm{Dist}&\left( \frac{1}{N}\sum_{n=1}^N \frac{\partial \ell(x_{n}^{\prime},y_{n}^{\prime})}{\partial W} \right. \\
%  & \left. - \nabla W\right) + \alpha \cdot \mathcal{R},
% \end{aligned}
% \end{equation}
\end{definition}

\textbf{2) GIA-L:} The fundamental concept behind GIA-L mirrors that of GIA-O, albeit with a distinction: GIA-L involves the initialization and optimization of $N$ pairs of latent vectors $\mathbf{z}^{\prime}$ and labels $\mathbf{y}^{\prime}$, reconstructing the private data by a generative adversarial network (GAN) \cite{goodfellow2014generative} denoted as \textit{G}.
\begin{definition}[GIA with Latent Space Optimization]
\begin{equation}
\label{eq2}
\begin{aligned}
&\{G(\mathbf{z}_{n}^{\prime*}), \mathbf{y}_{n}^{\prime*}\}_{n=1}^N =  \\
& \operatorname*{\text{argmin}}_{\{\mathbf{z}_{n}^{\prime}, \mathbf{y}_{n}^{\prime}\}_{n=1}^N} \mathrm{Dist}\left( \frac{1}{N}\sum_{n=1}^N \frac{\partial \ell(G(\mathbf{z}_{n}^{\prime}),\mathbf{y}_{n}^{\prime})}{\partial W} - \nabla W \right) + \alpha \mathcal{R}.
\end{aligned}
\end{equation}
% \begin{equation}
% \label{eq2}
% \begin{aligned}
% &\{G(z_{n}^{\prime*}), y_{n}^{\prime*}\}_{n=1}^N   \\
% &= \operatorname*{\text{arg min}}_{\{z_{n}^{\prime}, y_{n}^{\prime}\}_{n=1}^N} \mathrm{Dist}&\left( \frac{1}{N}\sum_{n=1}^N \frac{\partial \ell(G(z_{n}^{\prime}),y_{n}^{\prime})}{\partial W} \right. \\
% &\left.- \nabla W \right) + \alpha \cdot \mathcal{R}.
% \end{aligned}
% \end{equation}
\end{definition}
\par
\textbf{Analytic-based GIA} (GIA-A) aims to reconstruct training data precisely by formulating and solving equation systems that relate gradients to inputs. Early studies solve this problem by recursively inferring feature maps layer by layer from the gradients until the input is reconstructed~\cite{fan2020rethinking,zhu2020r}. However, these methods are limited to shallow, fully connected networks or the reconstruction of a single image. To improve the effectiveness, subsequent studies assume a more powerful malicious server capable of either \textbf{crafting}~\cite{fowl2021robbing,fowl2022decepticons,zhao2023resource} or \textbf{modifying}~\cite{boenisch2023curious,wen2022fishing} model parameters:

\textbf{1) GIA-A with malicious parameter crafting:} This type of method reconstructs the inputs by crafting and inserting malicious structures into the benign model. Fowl et al.~\cite{fowl2021robbing} introduce the ``\textit{imprint module}'', a huge linear layer inserted at the model's front maliciously. Through specialized weight initialization, this module enables the reconstruction of inputs exhibiting target properties (e.g., specific image brightness) from its gradients. This concept is later extended to language transformers. Fowl et al.~\cite{fowl2022decepticons} first disable all the attention layers and most of the feed-forward layers, and then \textit{insert the imprint modules into the feed-forward layers to separate and reconstruct tokens}. Zhao et al.~\cite {zhao2023resource} further proposed a \textit{sparsified imprint module} to mitigate the computational overhead for such methods.

\textbf{2) GIA-A with malicious parameter modifying:} This type of method does not insert any new structure but manually modifies a wide range of model parameters. Boenisch et al.~\cite{boenisch2023curious} reconstruct the inputs by \textit{modifying the parameters of the first fully-connected layer}. Specifically, they fine-tuned the layer's weights $W$ using an auxiliary dataset with the same distribution as the client's private data. This modification ensures that a sample $i$ in a batch activates only the neuron corresponding to row $W_i$, enabling the separation and reconstruction of batched samples. However, this method is limited to simple fully connected networks with ReLU activation functions. Wen et al.~\cite{wen2022fishing} further refined this technique by \textit{modifying the parameters in the classification layer} associated with the target class, setting all other parameters to zero. This modification ensures that the batched gradients reflect only the gradient of the target sample.

\textbf{Limitations of GIA-A in practice:} Currently, \textit{GIA-As face significant challenges in balancing utility and stealth in practical FL systems}. Without engaging in malicious behavior, an honest-but-curious server can only reconstruct a single image using shallow models through analytical methods~\cite{fan2020rethinking,zhu2020r}, posing limited threats in practical settings. While with malicious behaviors, the attacks could be easily detectable by clients. Such behaviors typically involve inserting highly unusual structures into the model or making large-scale parameter modifications, rendering the malicious model highly conspicuous to clients~\cite{geiping2020inverting,boenisch2023curious,wen2022fishing}. Moreover, recent studies have shown that the anomalous functionality of these malicious models can also be further detected in the gradient space~\cite{garov2023hiding}.\par%It was not until Folw et al. \cite{fowl2021robbing} made a breakthrough by assuming that a \textit{malicious} server could actively add a huge fully-connected layer to the global model, so that the input can be obtained directly by solving equations. From then on, follow-up works mainly focus on crafting\cite{chu2022panning,pasquini2022eluding} or modifying\cite{fowl2022decepticons,boenisch2023curious,wen2022fishing,zhao2023resource} the model parameters to inverse solve out the input.\par
\textit{(2) Modality.} As demonstrated in Fig.~\ref{fig:milestones}, most of the GIAs focus on Computer Vision (CV) tasks. Recent efforts have started investigating GIAs on Natural Language Processing (NLP) \cite{deng2021tag,balunovic2022lamp,gupta2022recovering} tasks. However, GIAs currently poses a limited threat on language models, and we make a further discussion in Sec.\ref{sec:dis}. Moreover, a recent study by Vero et al. in\cite{vero2022data} has explored GIA's applicability on tabular data.\par

\subsection{Defense} 
Cryptographic methods, such as secure multi-party computation\cite{mohassel2017secureml,bonawitz2017practical} and homomorphic encryption\cite{cheng2021secureboost}, have been applied in various privacy-preserving tasks. However, in FL systems, these techniques often result in significant computation and/or communication overheads\cite{wang2022protect} (disscussed in Appx.\ref{appx:sup}). Consequently, recent research efforts mainly opt for either \textbf{perturbing representations}\cite{sun2021soteria,scheliga2022precode} or \textbf{employing post-processing of gradients}\cite{li2022auditing,zhu2019deep} to defend against GIAs. \textbf{Representation perturbation} relies on the premise that if the representations during the forward propagation process are perturbed, the gradient would struggle to accurately convey features about the input. Several approaches, such as pruning\cite{sun2021soteria} or the integration of a variational module\cite{scheliga2022precode}, have been employed to implement such perturbations. However, these methods have demonstrated less effectiveness against current GIAs\cite{yue2023gradient}. Another defense approach is \textbf{post-processing on gradients}. Techniques like compression\cite{li2022auditing} and sparsification\cite{yue2023gradient} mislead adversaries by perturbing the gradients. These post-processing methods are commonly adopted in practical FL systems and exhibit potential in countering GIAs. \par
\underline{\textit{Our focus:}} To explore the real threat of GIAs in practical FL systems, our study assumes that the adversary is an \textbf{honest-but-curious} server. We focus on \textbf{optimization-based GIAs} for \textbf{image reconstruction} tasks because of its applicability and widespread interests. In addition, we consider \textbf{gradient post-processing} techniques as defensive methods due to their common applications in practice.\par
  %\underline{\textit{Our focus: }} In this work, we focus on \textbf{optimization-based} gradient inversion with \textbf{honest-but-curious} server in \textbf{image reconstruction}, considering that it poses a more general, practical and stealthy threat, and has received a lot of attention.\par
\section{Rethinking GIA in Practical FL}
\label{rethinking}
In this section, we identify three fundamental aspects affecting GIA. By examining the gap between the literature and practice, we highlight three key research questions (RQs) to reveal the potential threats posed by GIA in practical FL.\par

A practical FL system is a distributed machine learning framework where clients locally train models using \textit{several private samples} over \textit{multiple iterations} (e.g., mini-batch SGD). After local training, the clients share \textit{parameter updates} with a central server, which aggregates these updates to produce a new global model. This process repeats over multiple rounds until the convergence.

Therefore, let us go back to the drawing board to identify the fundamental aspects that affect GIAs in practical FL systems. Since GIAs reconstruct local data from shared updates, the critical step lies in understanding how these updates are generated and shared. First, \textbf{the local training setup serves as the foundation}, which determines how private data participate in training. Second,\textbf{ the model maps the inputs to parameter update}, through forward and backpropagation. Finally, \textbf{the raw update often undergoes post-processing before being shared}, to meet the security and efficiency requirements in deployments. For example, updates may be compressed to enhance communication efficiency.

\begin{figure}[t]
%\vspace{-2mm}
    \centering
    \includegraphics[width=0.47\textwidth]{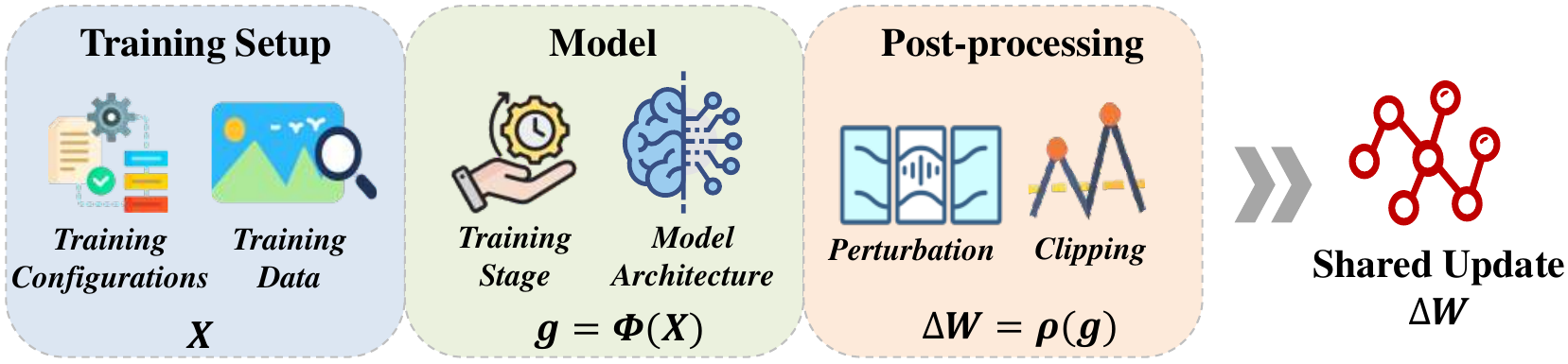}
    \caption{Three Fundamental Aspects of GIA.}
    \label{fig:rethinking}
    \vspace{-3mm}
\end{figure}

As illustrated in Fig.~\ref{fig:rethinking}, the transition from training data to the shared update follows these three critical steps. Consequently, evaluating the risk of GIA requires careful consideration of these aspects.
\textbf{Training setup.} The literature often assumes that the victim combines all its local data into a single batch ($B=N$) and updates $W^{g}_{t}$  for only one step ($U=1$), then share the gradient $\nabla W^{g}_{t}$, which benefits the adversary a lot. However, in practice, the client updates $W^{g}_{t}$ with mini-batch SGD, where $W^{k}_{t+1}$ are shared after training $W^{g}_{t}$ for multiple mini-batch steps. Therefore, the adversary could only obtain the update $W^{k}_{t+1}-W^{g}_{t}$. Besides, the training data are quite different in practice. First, the practical data dimensions (e.g., batches of $64$ images with the resolution of $128\times128$) are much larger than those in the literature (e.g., reconstruct $8$ images with the resolution of $32\times32$). Also, even if the data are of the same dimension, do they have the same risk of privacy leakage given the diversity of content? Based on these discussions, we aim to answer: 

\textbf{$\bullet$ RQ1:} How do the training setups affect GIAs in practical FL systems? (Sec.~\ref{sec:training})\par
 \label{rq1}
\textbf{Model.} The model plays a crucial role in mapping training data to gradient, but the literature treats it as a black box and underestimates its significance. In pursuit of performance, GIAs are often evaluated on specific architectures (e.g., ResNet-50 \cite{yin2021see,hatamizadeh2022gradvit} pre-trained with MOCO-V2\cite{chen2020improved}) or models with explicit initialization (a wide range of values from a uniform distribution\cite{zhu2019deep,zhao2020idlg,geiping2020inverting,jeon2021gradient}). In practice, an adversary may launch a GIA against any model, in any training stage, rather than a tailored one. Therefore, uncovering the model's black box and investigating its vulnerability is important to expose GIA's privacy threats:

\textbf{$\bullet$ RQ2:} What are the factors that influence the model's vulnerability to GIA? (Sec.~\ref{sec:model})\par
 \label{rq2}
 \textbf{Post-Processing.} Literature often assumes that the adversary could obtain the raw gradients directly. However, in practical FL systems, clients often perform post-processing on gradients before sharing them. For instance, gradients are commonly quantized to alleviate communication overhead\cite{lin2017deep}. Essentially, post-processing induces gradient drifting and potentially disrupts the adversary, which reminds us of an attractive trade-off problem:
 
\textbf{$\bullet$ RQ3:} Can FL systems naturally defend against GIAs with post-processing techniques, while ensuring the utility? (Sec.~\ref{sec:post-processing})\par

\section{Evaluation on Training Setup}
\label{sec:training}
The client's training setup, specifying what and how training data is utilized to compute the gradient, fundamentally affects the difficulty of reconstruction in practical FL systems. In this section, we investigate how the client's training setups affect GIA from two critical aspects: \textit{\textbf{training configurations}} and \textbf{\textit{training data}}. 

\subsection{Training Configurations}
\label{sec:training-1}
The client's training configurations depict how it organizes the data for training. The local dataset, consisting of $N$ samples, is divided into $\frac{N}{B}$ mini-batches. After completing $E$ epochs, the client shares the model $W_{E,0}$ following a total of $U = E \times \frac{N}{B}$ updates. Previous studies have often assumed ideal conditions ($B=N$, $E=1$) to maximize the reconstruction quality of N samples, where the adversary has access to the exact gradient of full batch data with a single update. However, in practical FL systems, an adversary can only perform GIA by approximating $(\mathbf{W}_{E,0}-\mathbf{W}_{0,0})/\eta$ where multiple gradients are squeezed into one update. In this subsection, we first analyze the impact of multiple updates on GIA in full-batch settings and then investigate the effectiveness of GIA against more general mini-batch SGD.

\subsubsection{Evaluate GIA against Full-Batch Updates}
We begin by examining the impact of multiple updates on GIA in the full-batch setting, where the victim client configures $B = N$ and uploads the updated model after $U$ local updates. First, we formalize the relationship between gradient and inputs and derive the reconstruction error introduced by multiple local updates. We then empirically evaluate the performance of GIA under varying numbers of local updates.   

% We consider a scenario where the client holds $N$ data samples and sets $B = N$. The client uploads $\mathbf{W_{U}}$ to the server after $U$ updates to $\mathbf{W_{0}}$. We start with a theoretical analysis of GIA against multiple local updates with a binary classification problem and then provide empirical evaluations on practical classification tasks.

\begin{figure}[t]
\vspace{-2mm}
    \centering
    \includegraphics[width=0.35\textwidth]{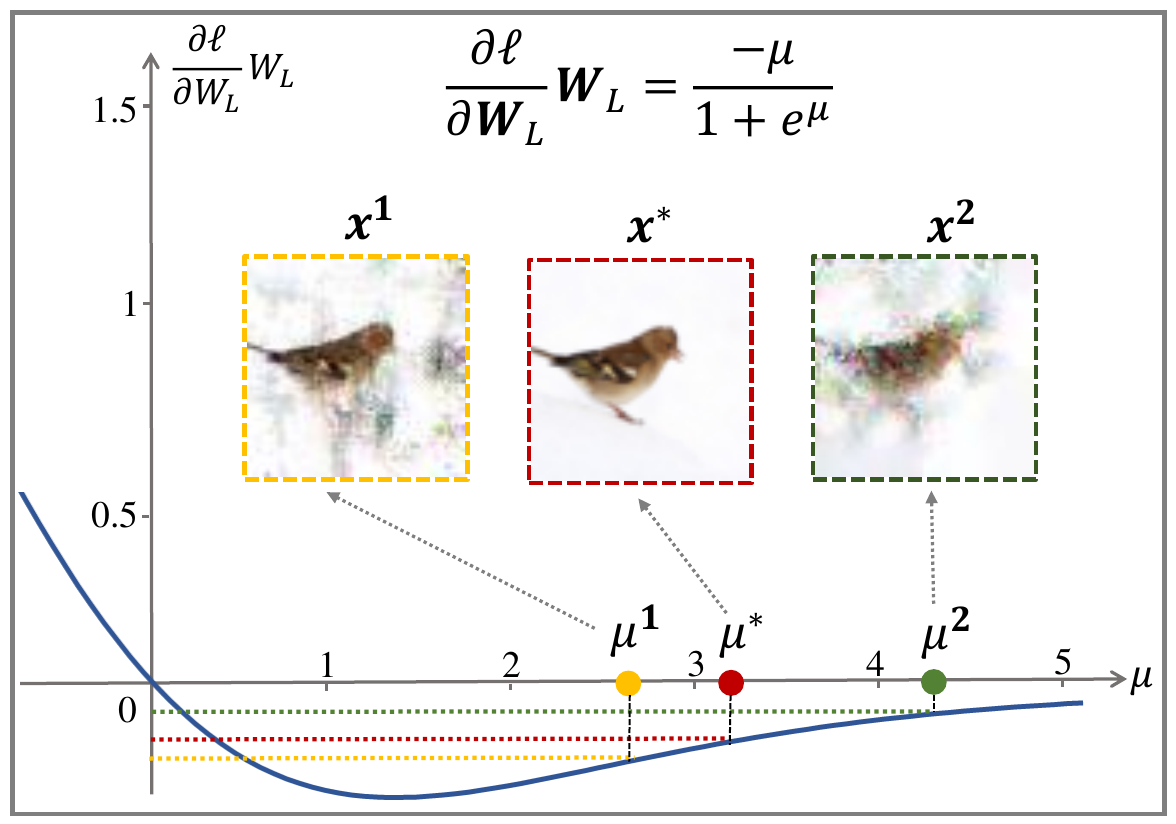}
    
    %\caption{Obfuscated $\mu$ Troubles the Reconstruction of \textbf{x}. \textcolor{red}{\textbf{Red dot ($\mu_{0}$)}} represents the the ground truth. \textcolor{blue}{\textbf{Blue dot ($\mu_{1}$)}} and \textcolor{green}{\textbf{green dot ($\mu_{2}$)}} represent the obfuscated $\mu$.}
    \caption{Dependence between Gradient $\frac{\partial\ell}{\partial \mathbf{W}}$, $\mu$ and Input $\mathbf{x}$. The ground-truth gradient corresponds to $\mu^{*}$, $\mathbf{x}^{*}$. When the gradients are obfuscated, they correspond to the inaccurate $\mu^{1}$, $\mu^{2}$, and $\mathbf{x}^{1}$, $\mathbf{x}^{2}$.}
    \label{fig:muandx}
    \vspace{-3mm}
\end{figure}

\textbf{Theoretical Analysis.} Consider a binary classification task ($y \in \{-1, 1\}$), and an $L$-layer fully connected network with activation function $\sigma$:
\begin{equation}
\label{eq:mu}
    \begin{aligned}
&\mu=y\mathbf{W}_{L}\mathcal{F}_{L-1}, \\
\end{aligned}
\end{equation}
\begin{equation}
\label{eq:f_l}
    \begin{aligned}
&\mathcal{F}_{L-1}=\sigma(\mathbf{W}_{L-1}\mathcal{F}_{L-2}) \quad \mathcal{F}_{L-2}=\sigma(\mathbf{W}_{L-2}\mathcal{P}(\mathbf{x})),
\end{aligned}
\end{equation}
where $\mu$ represents the logit, $\mathbf{W}_{L}$ denotes the augmented parameter matrix (including weight and bias) of the $L_{th}$ layer, $\mathcal{P}$ represents all layers previous to $L-2$, and $\mathbf{x}$ denotes the flattened vector of input data. Given the network's logit $\mu$, the loss function can be expressed as:
\begin{equation}
\label{eq:loss}
    \begin{aligned}
\ell=\log(1+e^{-\mu}).
\end{aligned}
\end{equation}
\begin{lemma}
\label{lemma:gdx}
For a fully connect network, the input $\mathbf{x}$ can be iteratively derived from gradient (Eq.~\eqref{eq:mux}) by first solving the logit $\mu$ (Eq.~\eqref{eq:gdmu}):
\begin{equation}\label{eq:gdmu}
\frac{\partial\ell}{\partial \mathbf{W}_{L}}\cdot \mathbf{W}_{L}=\frac{-\mu}{1+e^{\mu}},
\end{equation}
\begin{equation}\label{eq:mux}
\mathbf{x}=y\mu\mathbf{W}_L^\top\prod_{l=1}^{L-1}\mathbf{W}_l^\top\odot\sigma^{-1}.
\end{equation}
\end{lemma}
Lemma.~\ref{lemma:gdx} establishes a connection between the input $\mathbf{x}$ and the gradient by logit. If only one local update is performed, the adversary has the chance to reconstruct $x^*$ precisely, as shown in Fig.~\ref{fig:muandx}. However, if the client performs multiple updates, we can further deduce the reconstruction error that it imposes on the adversary:
\begin{theorem}
\label{the:updates}
Suppose the client updates the initial model $\mathbf{W}^{0}$ with the ground truth $\mathbf{x}^*$ 
for $U$ times to obtains $\mathbf{W}^{U}$. The adversary performs GIA by approximating $\frac{\mathbf{W}^{U} - \mathbf{W}^{0}}{\eta}$ and obtains the reconstructed $\mathbf{x}^{rec}$. The reconstruction error can be denoted as:
\begin{equation}
\label{eq:error}
\begin{aligned}
\mathbf{x}^\mathrm{rec}-\mathbf{x}^*=y(\mu^{rec}-\mu^*)\mathbf{W}_L^\top\prod_{l=1}^{L-1}\mathbf{W}_l^\top\odot\sigma^{-1},
\end{aligned}
\end{equation}
where $\mu^{rec}-\mu^*$ can be approximated by:
\begin{equation}
\label{eq:mudiff}
\begin{aligned}
\mu^{rec}-\mu^{*}\approx\frac{(e^{{\mu^{*}}}+1)^{2}\Sigma_{u=1}^{U-1}\frac{\partial\ell}{\partial \mathbf{W}_{L}^{u}}\mathbf{W}^0_{L}}{\mu^{*}e^{{\mu^{*}}}-e^{{\mu^{*}}}-1}.
\end{aligned}
\end{equation}
\end{theorem}
A detailed proof is provided in  Appx.~\ref{appx:proof}.\par
Theorem.~\ref{the:updates} reveals the obfuscation caused by multiple updates for an adversary to reconstruct input $\mathbf{x}$. Specifically, if $U>1$, the adversary can only obtain the update $\frac{\mathbf{W}^{U} - \mathbf{W}^{0}}{\eta}$ which consists of the precise gradient $\frac{\partial\ell}{\partial \mathbf{W}^{0}}$ and $U-1$  redundant terms $\sum_{u=1}^{U-1} \frac{\partial\ell}{\partial \mathbf{W}^{u}}$. Thus, the reconstructed $\mathbf{x}^{rec}$ will drift from $\mathbf{x}^*$, as the examples ($\mathbf{x}^1$ and $\mathbf{x}^2$) in Fig.~\ref{fig:muandx} show. Further, Eq.~\eqref{eq:error} and~\eqref{eq:mudiff} show that the reconstruction error gradually accumulates and obfuscates the GIA as $U$ increases.

\textbf{Empirical Analysis.} To empirically validate our theoretical conclusion, we evaluate the performance of the two SOTA GIAs (GIA-O with $p_{TV}$ and GIA-L with pretrained  DcGAN\cite{radford2015unsupervised}) as the number of local updates ($U$) increases on the datasets CIFAR10 (C10, $N=B=4$) and CIFAR100 (C100, $N=B=8$). We use the metric \textit{Learned Perceptual Image Patch Similarity (\textit{LPIPS}})\cite{zhang2018unreasonable}  to measure GIA's performance, where a smaller value represents higher reconstruction quality. The detailed setup is described in Appx.~\ref{appx:setup}. Tab.~\ref{tab:update} demonstrates that when $U$ is 1 or 2, the adversary can still reconstruct the data. However, as $U$ increases, the \textit{LPIPS} value exceeds $0.1$,  indicating that the reconstructed images are nearly unrecognizable. Furthermore, the reconstruction quality deteriorates as $U$ increases, which indicates that the redundant terms in Eq.~\eqref{eq:mudiff} gradually accumulate, and inaccuracy in the adversary's update approximation grows.
\begin{table}[t]
    \vspace{-2mm}
    \centering
    \renewcommand{\arraystretch}{0.5}
    \setlength\tabcolsep{3.38pt}
    \caption{GIA against Local Updates (\textbf{LPIPS}$\downarrow$). \textbf{Bold text} represents LPIPS value exceeds 0.1, indicating that the reconstructed image is completely unrecognizable.}
    \label{tab:update}%
    %\vspace{-4mm}
    \small % 这里可以替换为 \footnotesize, \scriptsize 等
    \begin{tabular}{ccccccc}
    \toprule
    \multirow{2}[4]{*}{\textbf{Dataset}} & \multirow{2}[4]{*}{\textbf{GIA}} & \multicolumn{5}{c}{\textbf{Number of Update ($U$)}} \\
\cmidrule{3-7}          &       & 1     & 2     & 4     & 6     & 8 \\
    \midrule
    \multirow{2}[4]{*}{CIFAR10} & O & 0.0189 & \textbf{0.1231} & \textbf{0.1499} & \textbf{0.1921} & \textbf{0.2522} \\
\cmidrule{2-7}          & L & 0.075 & 0.0952 & \textbf{0.1554} & \textbf{0.1607} & \textbf{0.1984} \\
    \midrule
    \multirow{2}[4]{*}{CIFAR100} & O & 0.0049 & 0.0404 & \textbf{0.1165} & \textbf{0.1653} & \textbf{0.2268} \\
\cmidrule{2-7}          & L & 0.0288 & 0.063 & \textbf{0.1205} & \textbf{0.1514} & \textbf{0.1852} \\
    \bottomrule
    \end{tabular}%
\vspace{-3mm}
\end{table}%

\subsubsection{Evaluate GIA against Mini-Batch Updates}
Here, we explore a more practical scenario in which the client shares the $\textbf{W}_{U}$ after several local mini-batch SGD updates ($B<N, U=N/B$).  We examine two cases based on whether the adversary knows $B$. \textbf{Worst case (WST.):} Previous studies\cite{zhu2019deep,xu2022agic} often assume that the adversary has full knowledge of $B$ and other local training configurations. This knowledge allows the adversary to simulate the client's mini-batch updates more accurately, closely matching $\frac{\mathbf{W}_{U}-\mathbf{W}_{0}}{\eta}$, which maximizes the effectiveness of GIA. \textbf{Best case (BST.):} However, in practice, the client is not obligated to provide local training details to the server, thus the adversary can only approximate the update by conducting one-step SGD ($B=N, U=1$). This scenario significantly reduces the effectiveness of GIA because the adversary, lacking knowledge of the training configurations, cannot accurately approximate the update. 

We evaluate GIAs in both cases, as shown in Tab.~\ref{tab:mini}. We varied the number of mini-batches by adjusting the value of $N/B$ on the CIFAR10 ($B=4$) and CIFAR100 ($B=8$) datasets. The results indicate that the reconstruction fails as the number of mini-batches increases for both cases (indicated by the bold \textit{LPIPS} value exceeding $0.1$), even if the adversary has access to local training configurations. Additionally, it is important to note that the maximum number of updates evaluated in Tab.~\ref{tab:mini} is 10, which is significantly lower than what is typical in the practical FL system. \textit{Given that each client typically possesses several thousand to tens of thousands of data points, the number of local updates would be much higher, thereby posing an even greater challenge to GIA}.
\begin{table}[t]
%\vspace{-1mm}
    \centering
    \renewcommand{\arraystretch}{0.3} % 行间距更紧凑
    \setlength\tabcolsep{1.0pt} % 列间距更紧凑
    \footnotesize % 缩小字体
    \caption{GIA against Mini-Batch SGD (\textbf{LPIPS}$\downarrow$).}
    \label{tab:mini}%
    \begin{tabular}{cccccccccc}
    \toprule
    \multirow{2}[4]{*}{\textbf{Dataset}} & \multirow{2}[4]{*}{\textbf{GIA}} & \multirow{2}[4]{*}{\textbf{Case}} & \multicolumn{7}{c}{\textbf{Number of Mini-batch}} \\
\cmidrule{4-10}          &       &       & 1     & 2     & 3     & 4     & 6     & 8     & 10 \\
    \midrule
    \multirow{4}[8]{*}{C10} & \multirow{2}[4]{*}{O} & WST.  & \multirow{2}[4]{*}{0.0172} & 0.061 & \textbf{0.1064} & \textbf{0.1117} & \textbf{0.1456} & \textbf{0.1314} & \textbf{0.1426} \\
\cmidrule{3-3}\cmidrule{5-10}          &       & BST.  &       & 0.0641 & \textbf{0.1088} & \textbf{0.1075} & \textbf{0.1398} & \textbf{0.1403} & \textbf{0.1445} \\
\cmidrule{2-10}          & \multirow{2}[4]{*}{L} & WST.  & \multirow{2}[4]{*}{0.0669} & 0.0672 & 0.0834 & \textbf{0.1043} & \textbf{0.1436} & \textbf{0.1285} & \textbf{0.1376} \\
\cmidrule{3-3}\cmidrule{5-10}          &       & BST.  &       & 0.0833 & 0.0859 & \textbf{0.1017} & \textbf{0.1444} & \textbf{0.1318} & \textbf{0.1451} \\
    \midrule
    \multirow{4}[8]{*}{C100} & \multirow{2}[4]{*}{O} & WST.  & \multirow{2}[4]{*}{0.0045} & 0.0732 & 0.0792 & \textbf{0.1003} & \textbf{0.1063} & \textbf{0.1193} & \textbf{0.1273} \\
\cmidrule{3-3}\cmidrule{5-10}          &       & BST.  &       & 0.0581 & 0.0759 & 0.0928 & \textbf{0.1139} & \textbf{0.1183} & \textbf{0.1271} \\
\cmidrule{2-10}          & \multirow{2}[4]{*}{L} & WST.  & \multirow{2}[4]{*}{0.0436} & 0.0725 & 0.0788 & 0.0931 & \textbf{0.1017} & \textbf{0.1100} & \textbf{0.1174} \\
\cmidrule{3-3}\cmidrule{5-10}          &       & BST.  &       & 0.0776 & 0.0817 & 0.093 & \textbf{0.1016} & \textbf{0.1076} & \textbf{0.1178} \\
    \bottomrule
    \end{tabular}%
\vspace{-3mm}
\end{table}%
\begin{mdframed}[backgroundcolor=gray!10, roundcorner=5pt]
\textbf{(Insight \ref{sec:training-1})} The training configurations significantly impact GIA's effectiveness. Specifically, as the number of local updates increases, the reconstruction becomes increasingly challenging and even fails.%, particularly without insight into the client's training details, such as batch size ($B$).
%\neil{A reviewer can argue that server may know the batch size since the algorithm is provided by the server. Removing "particularly without insight ...such as batch size..'' does not influence this take-away message, but makes it less likely to be challenged by a reviewer.}
\end{mdframed}
\subsection{Training Data}
\label{sec:training-2}
The primary goal of GIAs is reconstructing training data. To investigate its impact, we evaluate the performance of GIAs on two basic data properties, \textbf{dimension} and \textbf{content}. To explore the impact of data dimension, we first give a theoretical upper bound on whether the adversary can accurately reconstruct the data as dimension grows, and then empirically evaluate the SOTA GIAs across various resolutions and batch sizes in practical FL. Subsequently, we investigate the influence of image content on GIAs—an interesting but largely overlooked aspect in existing literature. 
\subsubsection{Data Reconstruction across Wide Dimensions}
\label{sec:training-2-1}
Intuitively, reconstruction becomes more difficult with higher data dimensions (larger batch sizes, higher resolution). This raises a critical question: \textbf{is there a theoretical upper bound that makes it impossible for an adversary to accurately reconstruct data beyond a certain dimension?}
\begin{theorem}
\label{the:bound}
Suppose the ``input-gradient" function $\frac{\partial\ell}{\partial \mathbf{W}} = \Phi\left(\mathbf{x}\right)$, where the gradient $\frac{\partial\ell}{\partial \mathbf{W}} \in \mathbb{R}^{1 \times p}$, the input data $\mathbf{x} \in \mathbb{R}^{B \times D}$, $B$ and $D$ are batch size and data size. If $p<B \times D$, then there exists at least one mask $\Delta$ such that, $\Phi\left(\mathbf{x}\right)=\Phi\left(\mathbf{x}+\Delta\right)$.
\end{theorem}
A detailed proof is provided in  Appx.~\ref{appx:proof}.\par
Theorem.~\ref{the:bound} suggests that when the data dimension exceeds the number of model parameters, there will always exist at least one fake data point $\mathbf{x}+\Delta$ that differs from the ground truth $\mathbf{x}$ but yields the same gradient. In other words, \textit{even if the adversary has the strongest ability to approximate the received gradient, it will still be unable to accurately reconstruct the ground truth}. In practice, a bottleneck emerges that hinders data reconstruction when the dimensionality surpasses a certain threshold. Due to the model's sparsity and the adversary's limited computational resources, this threshold will be substantially lower than the theoretical value, $p$.

To demonstrate how the increasing data dimension restricts the effectiveness of GIAs, we evaluate four GIAs across various settings, including batch sizes ranging from $1$ to $100$ and image resolutions from $32\times32$ to $512\times512$. We evaluate the baseline $G_O$ (GIA-O with gradient-matching loss only) and three SOTA GIAs: $G_O$+$p_{TV}$ (GIA-O with prior total variation\cite{geiping2020inverting}), $G_O$+$p_{BN}$ (GIA-O with prior BN statistics\cite{yin2021see}), and $G_L$ (GIA-L\cite{jeon2021gradient,li2022auditing}). We evaluate GIAs under FedSGD\cite{konevcny2015federated} using ResNet-18 on CIFAR10, CIFAR100, and ImageNet-1K\cite{deng2009imagenet} datasets. The performance of GIA is measured by \textit{LPIPS}. Detailed setup is provided in Appx.~\ref{appx:setup}.\par %Since it mainly searches in the Latent space, we denote this attack as GIA-L. Similarly, for GIAs with priors (e.g. $p_{TV}$ and $p_{BN}$), they search in the data (image) space and hence denoted as GIA-D.\par
\begin{figure}[t]
\centering
\vspace{-2mm}
\begin{subfigure}{0.45\linewidth}\centering\includegraphics[width=1\linewidth]{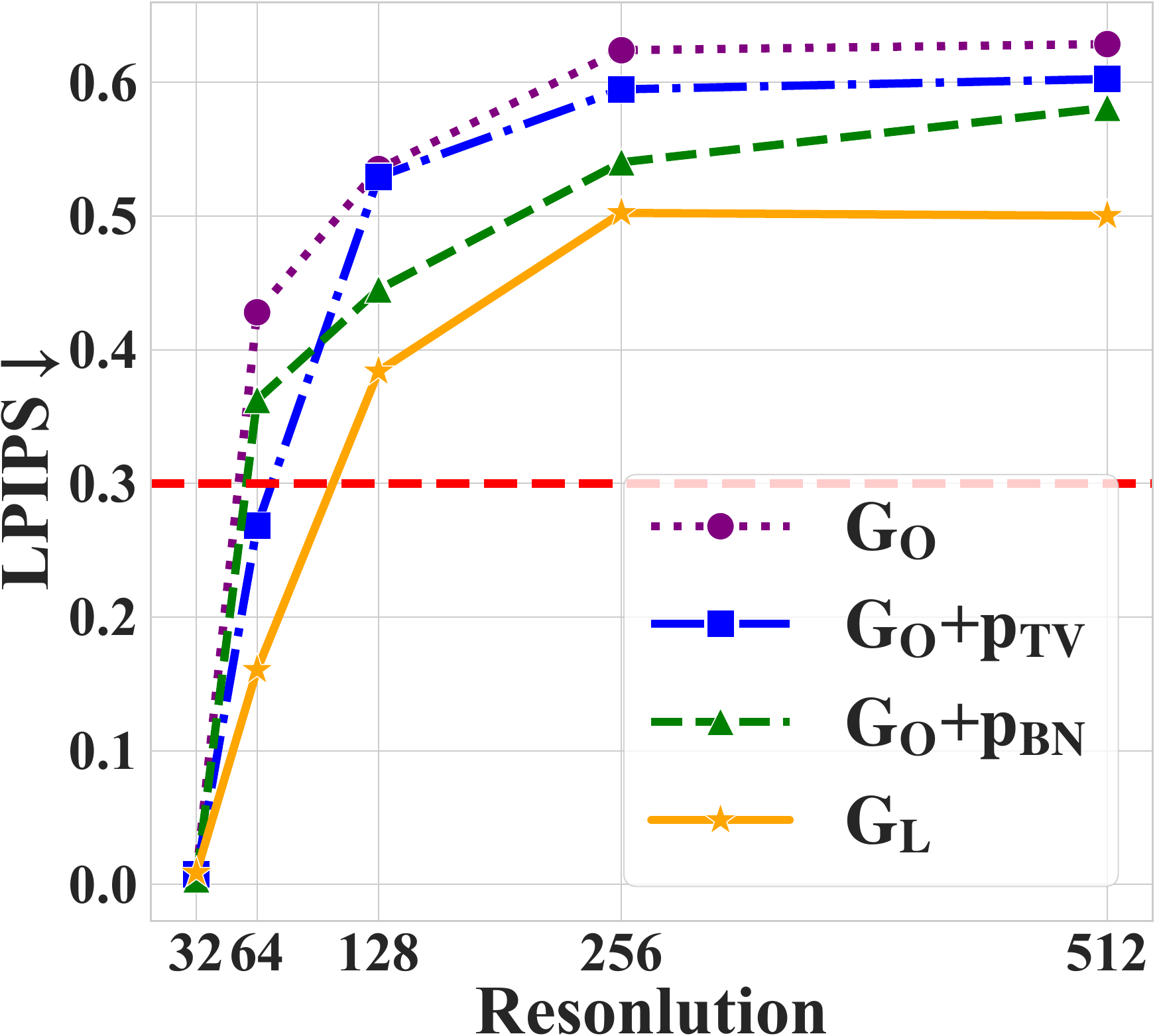}\caption{GIA on Resolutions}\label{pic:pixel}
\end{subfigure}
\centering
\begin{subfigure}{0.45\linewidth}\centering\includegraphics[width=1\linewidth]{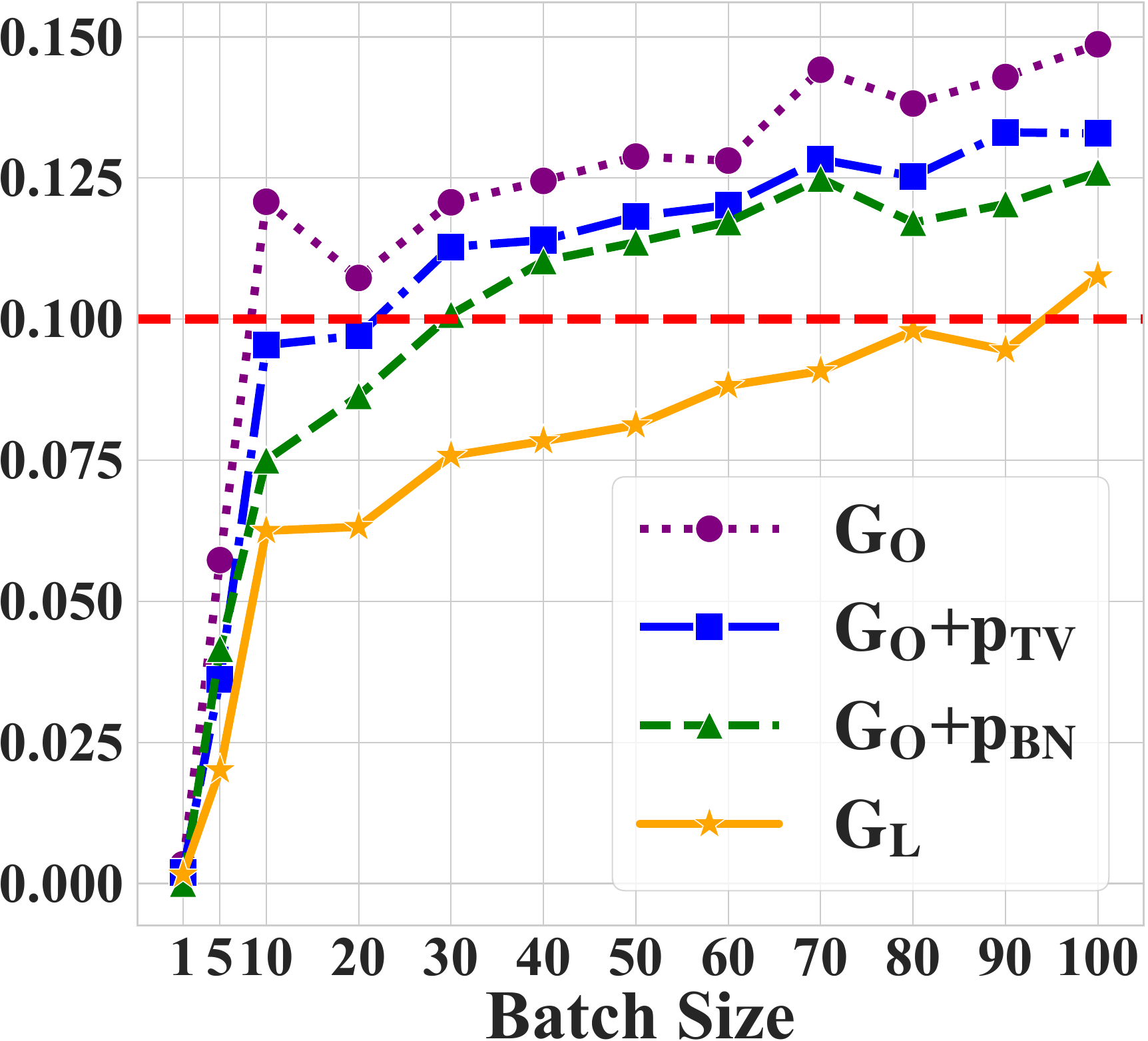}\caption{GIA on Batch Sizes}\label{pic:bs}
\end{subfigure}
\vspace{-2mm}
\caption{GIA on Series of Data Dimensions.}
\label{pic:capacity}
\vspace{-3mm}
\end{figure}
Fig.~\ref{pic:capacity} illustrates the statistical results, and LPIPS values above the red lines mean that the reconstructed images cannot disclose the privacy of the original images. Our findings indicate that (1) \textit{GIA performs well in reconstructing low-dimensional data} (resolution $<64\times64$ and batch size $<30$), but has difficulty in reconstructing high-dimension data. (2)\textit{ GIA-L presents a better performance in reconstructing high-dimensional data compared to GIA-O}. GIA-L benefits from two aspects: First, GIA-L has a smaller search space, which allows the optimizer to better find the optimal. Second, generative knowledge guarantees the fidelity. 

\begin{mdframed}[backgroundcolor=gray!10, roundcorner=5pt]
\textbf{(Insight \ref{sec:training-2-1})} An increase in data dimension weakens the effectiveness of GIA, and a bottleneck exists that prevents adversaries from effectively reconstructing higher-dimension data in practice.
\end{mdframed}

\subsubsection{Data Reconstruction across Various Contents}
\label{sec:training-2-2}
In practical scenarios, clients store diverse training data that share the same dimension and label space but hold various contents. Therefore,  \textbf{does GIA pose the same level of privacy risk for data with different contents?}

To explore the effectiveness of GIA against various data contents, we investigate two ``canary-testing'' categories. \textbf{Category 1: Data with semantic details.} In practice, the degree of data privacy leakage is often determined by the quality of reconstructed key semantic details, rather than overall similarity. \textbf{Category 2: Out-of-distribution (OOD) data}. This category is specific to GIA-L. Previous studies often assume that the adversary has access to a dataset with an identical-distribution (ID) to the client's data for generator pre-training. However, in practice, the adversary faces the OOD challenge because the client is not required to disclose its data distribution. As a result, the adversary must rely on public datasets for pre-training, leading to distributional bias.\par

\begin{figure}[t]
\vspace{-1mm}
    \centering
    \includegraphics[width=0.45\textwidth]{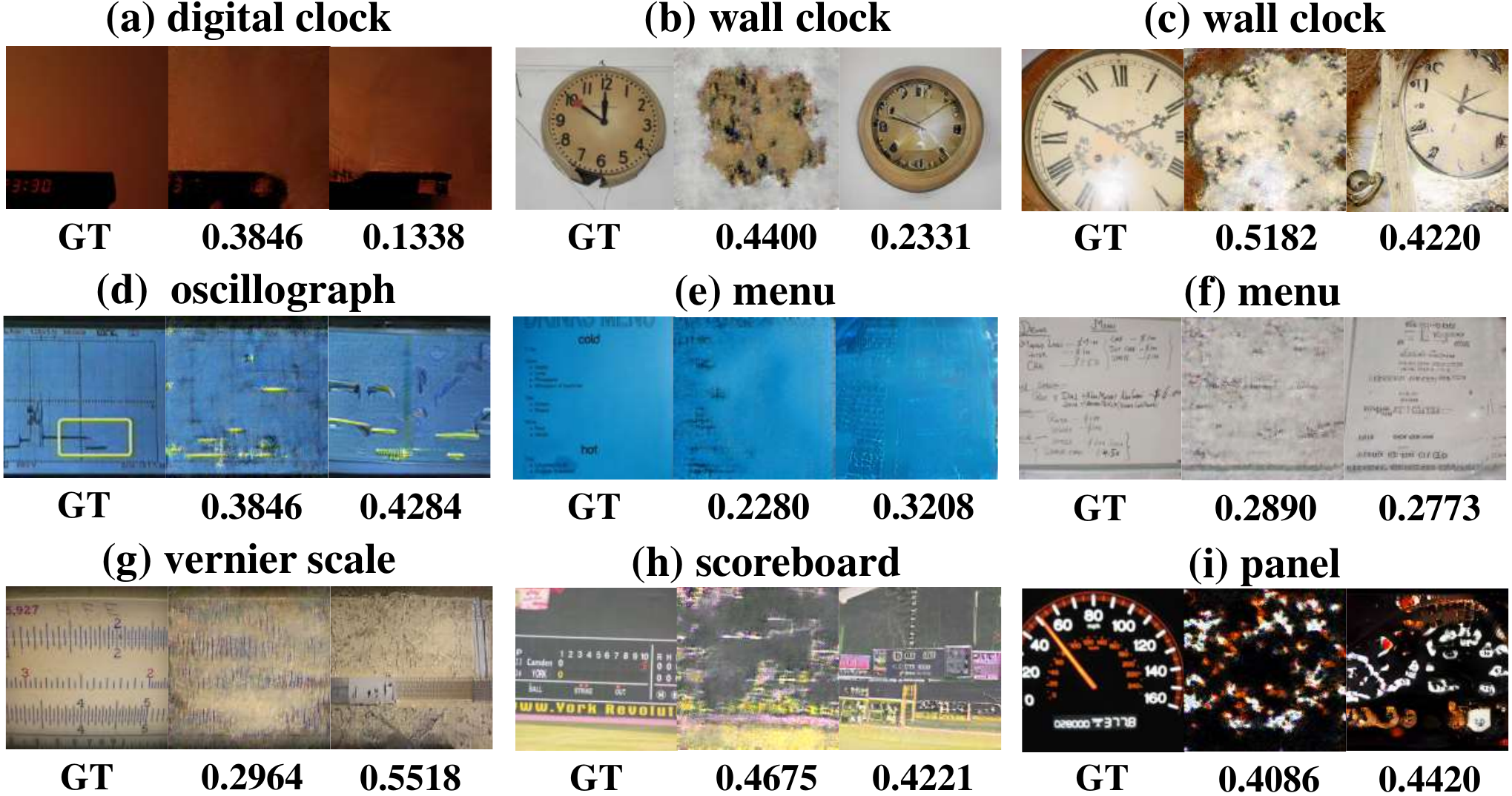}
    
    \caption{Failure to Reconstruct Semantic Details, Limited Privacy Leakage. (\textbf{Left:} Ground-Truth, \textbf{Middle:} Results of GIA-O, \textbf{Right:} Results of GIA-L, \textbf{LPIPS}$\downarrow$).}
    \label{fig:details}
    \vspace{-3mm}
\end{figure}

\textit{\textbf{\Rmnum{1}. Semantic details diminish the effectiveness of GIAs.}} To reveal the impact of semantic details on GIAs, we choose two types of attacks: GIA-O with priors including both $p_{TV}$ and $p_{BN}$, and GIA-L with BigGAN’s generator\cite{brock2018large} pre-trained on ImageNet-1K\cite{deng2009imagenet}. The experiments are conducted on ImageNet-1K dataset, and setup details are provided in the Appx.~\ref{appx:setup}. 
As shown in Fig.~\ref{fig:details}, the reconstructed images reveal very little private information because all the crucial details are lost. Specifically, Fig.~\ref{fig:details} \textit{(a)-(c)} show the results of reconstructing \textit{clock} images. Although the contours can be reconstructed, the crucial semantic details, such as the correct time, are completely missed. The neglect of semantic details is further exemplified in Fig.~\ref{fig:details} \textit{(d)-(f)}, where only the background can be reconstructed without recognizing any critical contents (e.g., numbers, text). In addition, we find that images with semantic details are much more challenging to fit and tend to mislead GIAs, as shown in Fig.~\ref{fig:details} \textit{(g)-(i)}. The reasons for GIA's negligence of semantic details stem from two aspects: (1) the gradient reflects more the capture of category features than semantic details; (2)  the pre-trained generator only learns class-wise features rather than identical details.\par

\textit{\textbf{\Rmnum{2}. OOD data impedes the generalization capability of GIA-L.}} To explore the effectiveness of GIA-L against OOD data, we assume that the adversary has access to ImageNet-1K\cite{hendrycks2021natural} dataset, and evaluate the GIAs on four OOD datasets. \textbf{Places} tests the performance of GIA-L in images involving complex background information\cite{huang2021mos}, \textbf{Textures}\cite{cimpoi2014describing} and \textbf{Objects}\cite{hendrycks2021natural} contain various pattern or object distributions, and \textbf{Styles} is designed for cross-style generalization\cite{li2017deeper}.

\begin{figure}[t]
\vspace{-2mm}
    \centering
    \includegraphics[width=0.45\textwidth]{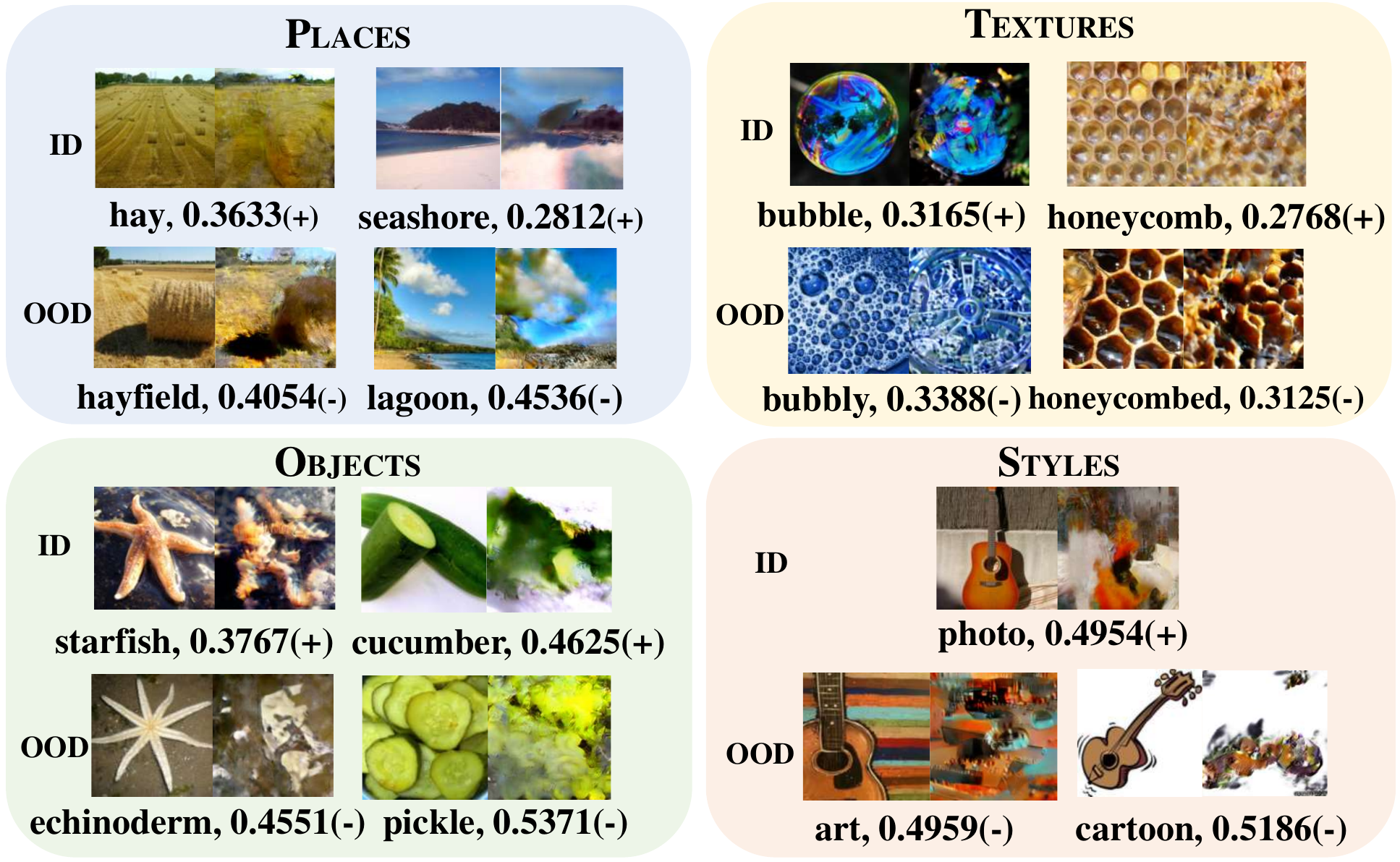}

    \caption{OOD Challenge for GIA-L. (\textbf{Left:} Ground-Truth, \textbf{Right:} Reconstructed Image,  \textbf{LPIPS}$\downarrow$. Each column is a pair of samples with \underline{\textbf{similar labels}}, \textbf{Up:} Sample from ImageNet-1K\cite{deng2009imagenet}(ID) dataset, \textbf{Bottom:} Sample from the OOD datasets).}
    \label{fig:OOD}
    \vspace{-4mm}
\end{figure}

Fig.~\ref{fig:OOD} demonstrates the reconstruction performance of GIA-L on ID and OOD data. The results indicate that GIA-L performs badly on OOD data due to its limited generative capability. \textbf{For each pair of samples labeled similarly, the LPIPS values for reconstructing ID images are much lower than those of reconstructing OOD ones}. In the case of Textures, GIA-L are much familiar with ``discrete and transparent" bubbles than ``dense and blue" ones, as demonstrated in Fig.~\ref{fig:OOD}, respectively. Another example involves different styles of guitars, where GIA-L lacks knowledge about the characteristics in \textit{art} or \textit{cartoon} style.

\begin{mdframed}[backgroundcolor=gray!10, roundcorner=5pt]
\textbf{(Insight \ref{sec:training-2-2})} The effectiveness of GIA varies against various data contents. Thus, GIA's privacy risks are often overestimated, particularly when reconstructing semantic details and OOD data.
\end{mdframed}

\section{Evaluation on Model}
\label{sec:model}
%Yin et al.\cite{yin2021see} performs their experiments on a ResNet-50 pre-trained with MOCO-V2\cite{chen2020improved} and a specific fine-tuningBesides, although Hatamizadeh, et al analyzed the effect of the training phase on gradient inversion in their recent work\cite{hatamizadeh2023gradient}, they only conducted experiments on ResNet-18 and only gave visual reconstruction results without giving further explanation and analysis.
The model determines the mapping from training data to the gradient. During the FL training phase, the adversary acquires the trained model and its shared gradient at a particular stage (round) for GIA. In this section, we investigate the impact of the model on gradient inversion from the perspectives of \textbf{stage} and \textbf{architecture}. To illustrate and quantify the model's vulnerability to GIAs at different stages, we introduce a novel \textbf{I}nput-\textbf{G}radient \textbf{S}moothness \textbf{A}nalysis (\textbf{IGSA}) method. Through IGSA, we empirically assess the resistance of nine models against GIAs at different stages during the whole FedAvg training phase. Subsequently, we explore the sensitivity of GIAs to model architecture.

\subsection{Training Stage}
\label{sec:model-1}
\subsubsection{Input-Gradient Smoothness Analysis (IGSA)}
The ``input-gradient" function, denoted as $\Phi(\cdot)$, which includes both forward and backward propagation, maps input data to gradients, forming the foundation of GIA's objective function. To understand how different training stages impact GIA, we begin by analyzing the properties of $\Phi(\cdot)$.

The smoothness of the objective function influences the difficulty in locating the global optimum, a concept that applies to GIA as well \cite{qu2017harnessing}. Specifically, smooth functions are easily optimized and vice versa, they tend to fall into local optima.

Therefore, we propose a novel method termed \textbf{IGSA} to characterize the model's resilience to GIAs:
\begin{equation}
\begin{aligned}
IGSA^\mathbf{X} & =\left(\mathbf{\mathbb{E}}_{\Delta x}\left[\left\|\Phi(\mathbf{X})-\Phi(\mathbf{X}+\Delta \mathbf{X}) \cdot \omega\right\|_{2}\right]\right)^{-1} \\
& =K\left({\sum_{k=1}^K\left\|\mathbf{J}\left(\mathbf{X_k}\right) \cdot \omega\right\|_{2}}\right)^{-1},
\end{aligned}
\end{equation}
where $\mathbf{J}(\mathbf{X})=\frac{\partial \Phi(\mathbf{X})}{\partial \mathbf{X}}$ denotes the Jacobi matrix to compute the first-order derivative of  $\Phi\left(\cdot\right)$, which reflects \textbf{how drastically the gradient changes in response to input perturbations}. To estimate the IGSA values, we utilize $K$ samples located within a radius $r$ around  $\mathbf{X}$. Considering that the gradients span across $L$ layers, we construct a vector $\omega= \left[1+1/L, 1+1/(L-1), \ldots, 2\right]$. This vector assigns higher weights to the shallow layers when averaging $l_{2}$-norms. \textbf{A higher IGSA value indicates that $\Phi\left(\cdot\right)$ possesses a greater degree of smoothness, enabling it easier for the adversary to locate and reconstruct training data.} \par
% \begin{figure}[t]
% %\vspace{-1mm}
% \centering
% \begin{subfigure}{0.4\linewidth}\centering\includegraphics[width=1\linewidth]{figs/smooth.pdf}\caption{Smooth Landscape}\label{pic:smooth}
% \end{subfigure}
% \centering
% \begin{subfigure}{0.4\linewidth}\centering\includegraphics[width=1\linewidth]{figs/oscillatory.pdf}\caption{Oscillatory Landscape}\label{pic:oscillatory}
% \end{subfigure}
% \caption{Examples of Objective Function's Landscape.}
% \label{pic:smoothness}
% \vspace{-4mm}
% \end{figure}
\subsubsection{Evaluating Model's Vulnerability to GIA in Different Stages}
In this part, we analyze and explain the model's vulnerability to GIAs during the training process. We choose $9$ models from four families that are widely used in vision tasks, i.e., VggNet-(11, 16, 19)\cite{simonyan2014very}, ResNet-(18, 50, 152), DenseNet121\cite{huang2017densely}, and InceptionNet (GoogleNet\cite{szegedy2015going} and Inception-v3\cite{szegedy2016rethinking}). \par
\begin{figure}[t]
%\vspace{-1mm}
    \centering
    \includegraphics[width=0.47\textwidth]{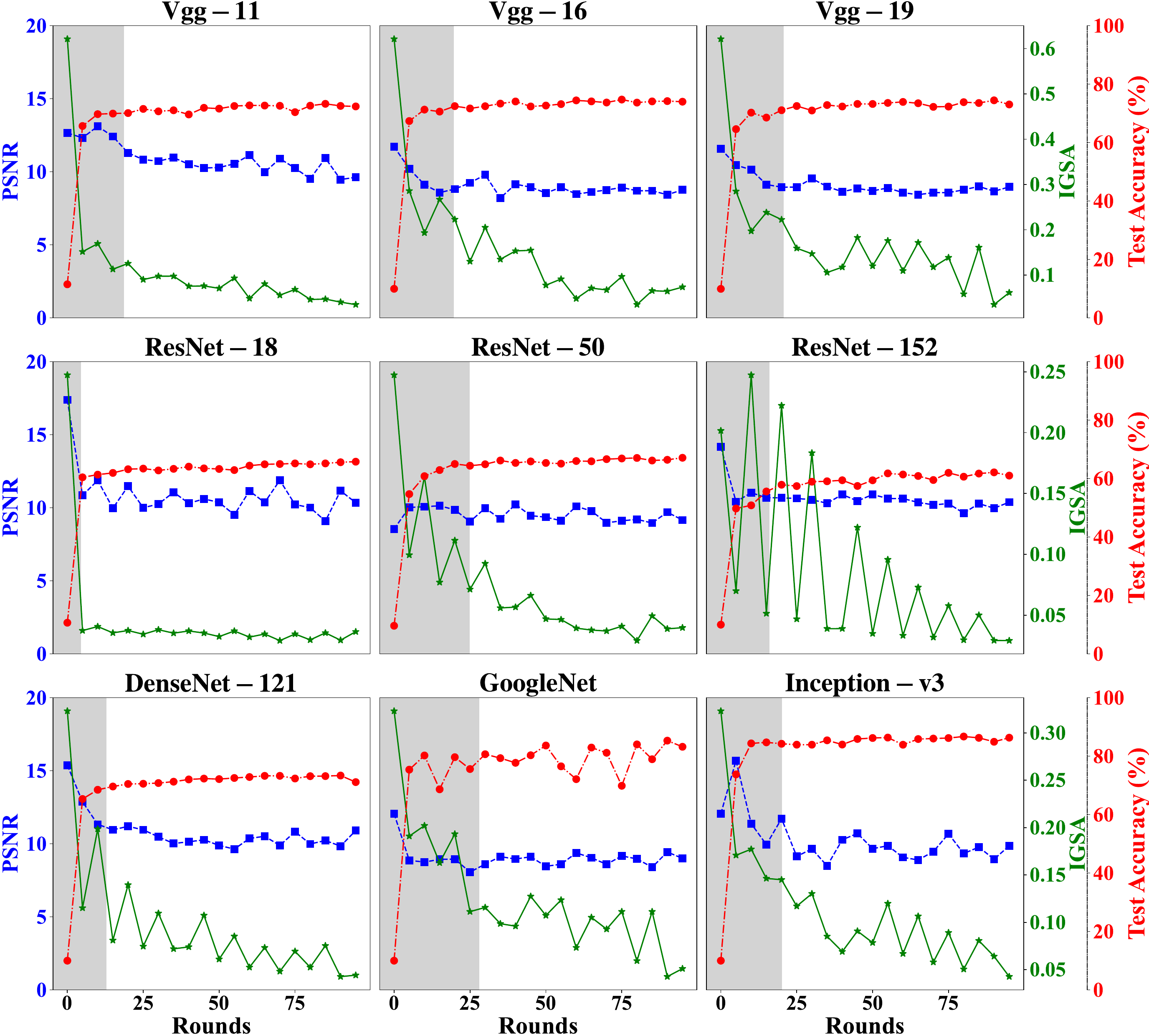}

    \caption{Model's Resistance to GIA during FL Training. Model's test accuracy: \textcolor{red}{\textbf{Test Accuracy}$\uparrow$}, reconstruction quality: \textcolor{blue}{\textbf{PSNR}$\uparrow$}, resistance to GIA: \textcolor{green}{\textbf{IGSA}$\downarrow$}. \textcolor{gray}{\textbf{Gray areas}} represent the rounds with high privacy risk.}
    \label{fig:IGSA}
    \vspace{-2mm}
\end{figure}

%the IGSA values for each round in the training process are normalized, as shown in the green curves for subplots in Fig.~\ref{fig:IGSA}, wher
%We evaluate GIAs under FedSGD\cite{konevcny2015federated} setting with ResNet-18 model on CIFAR10 and CIFAR100 datasets. The performance of GIA is measured by \textit{Learned Perceptual Image Patch Similarity (LPIPS)}\cite{zhang2018unreasonable}, whose smaller values indicate better reconstruction. Details of the setup are given in Appx.\ref{appx:setup}.\par
\textbf{Experimental setup:} We evaluate nine models using the CIFAR10 dataset. Each model undergoes $100$ training rounds. During each round, we launch GIA on the current model while assessing the accuracy and IGSA value. The GIA performance is measured by PSNR (Peak Signal-to-Noise Ratio). To enhance clarity, we average the results every five rounds in Fig.~\ref{fig:IGSA}. Further details are provided in Appx.~\ref{appx:setup}. \par 
%\textbf{Experimental setup: }We simulate a FedAvg setting in which an honest-but-curious server performs GIA on a certain client. Focusing on model, we use Geiping's algorithm\cite{geiping2020inverting} as a baseline GIA, controlling the auxiliary constrains as much as possible, and attacking only through gradients and a simple total variance prior (with weights of $1e^{-5}$). We train 200 rounds on the CIFAR10 dataset for 9 models under the FedAvg setup, with 5 clients. And batch size is set to 64, local epochs are 10. In each round, we record the accuracy of the current global model on the test set and perform GIA on the saved model while evaluating the similarity of the images with PSNR (Peak Signal-to-Noise Ratio). For IGSA, we randomly select $100$ different samples in the CIFAR10 averaged to fairly calculate the IGSA value for the each round. For each calculation, we set $K$ to $10000$ and radius $r$ to $1e^{-3}$. Besides, for each model, the IGSA values in the training process are normalized, as shown in the green curves for subplots in Fig.~\ref{fig:IGSA}.\par 
Fig.~\ref{fig:IGSA} illustrates the resilience of the nine models against the GIA during training. We can intuitively observe that \textbf{GIA poses higher risks in the early training stages}. For instance, in the case of the Vgg-11 model, initially, the PSNR value is high, but as the training process continues, it consistently stays below $10$ after approximately $15$ rounds, indicating the reconstructed images cause no privacy leakage. Moreover, as indicated by the test accuracy curves, we note that the model's resistance correlates well with the fitting degree. The model will be more fragile in the underfitting stages. Additionally, the IGSA curves exhibit distinct phases, where the reconstruction quality is acceptable when IGSA values remain high. Conversely, reconstruction tends to fail when IGSA curves drop and fluctuate at lower levels.\par 

\begin{mdframed}[backgroundcolor=gray!10, roundcorner=5pt]
\textbf{(Insight \ref{sec:model-1})} During FL training, early-stage models are more vulnerable to GIA, while late-stage ones have good resistance.
\end{mdframed}

\subsection{Model Architecture}
\label{sec:model-2}
The model architecture determines how features are extracted and the information flow, further influencing the path of backpropagation and gradient computation, thereby potentially affecting GIAs. In this subsection, we study the impact of model architecture on the GIA from two perspectives: \textbf{structures} and \textbf{micro designs}. % Therefore, by both empirical and theoretical analysis, we aim to find their rflowelationship with GIA.

\subsubsection{Model Structure: A Double-edged Sword}
\label{sec:model-2-1}
Model structure refers to the way layers are connected to each other. In this subsection, we explore the effect of model structures on the GIA with two widely used cases in practice: \textbf{skip connection}\cite{he2016deep} (used in models like ResNet and DenseNet families) and \textbf{net-in-net}\cite{szegedy2015going} (used in models like GoogleNet and other InceptionNet families).\par
\begin{figure}[t]
\vspace{-2mm}
\centering
\begin{subfigure}{0.47\linewidth}\centering\includegraphics[width=1\linewidth]{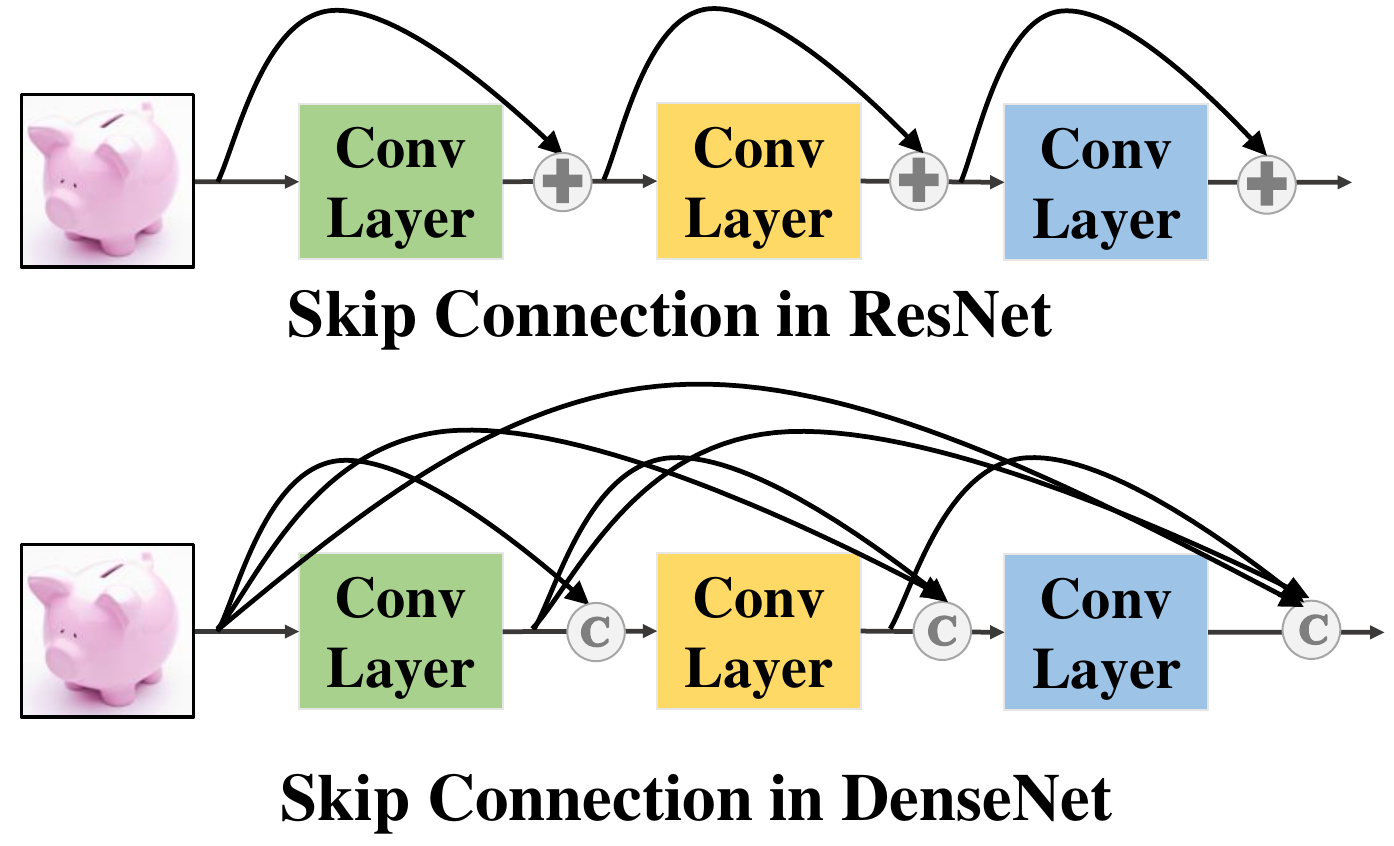}\caption{Skip Connection}\label{pic:res_connection}
\end{subfigure}
\centering
\begin{subfigure}{0.47\linewidth}\centering\includegraphics[width=1\linewidth]{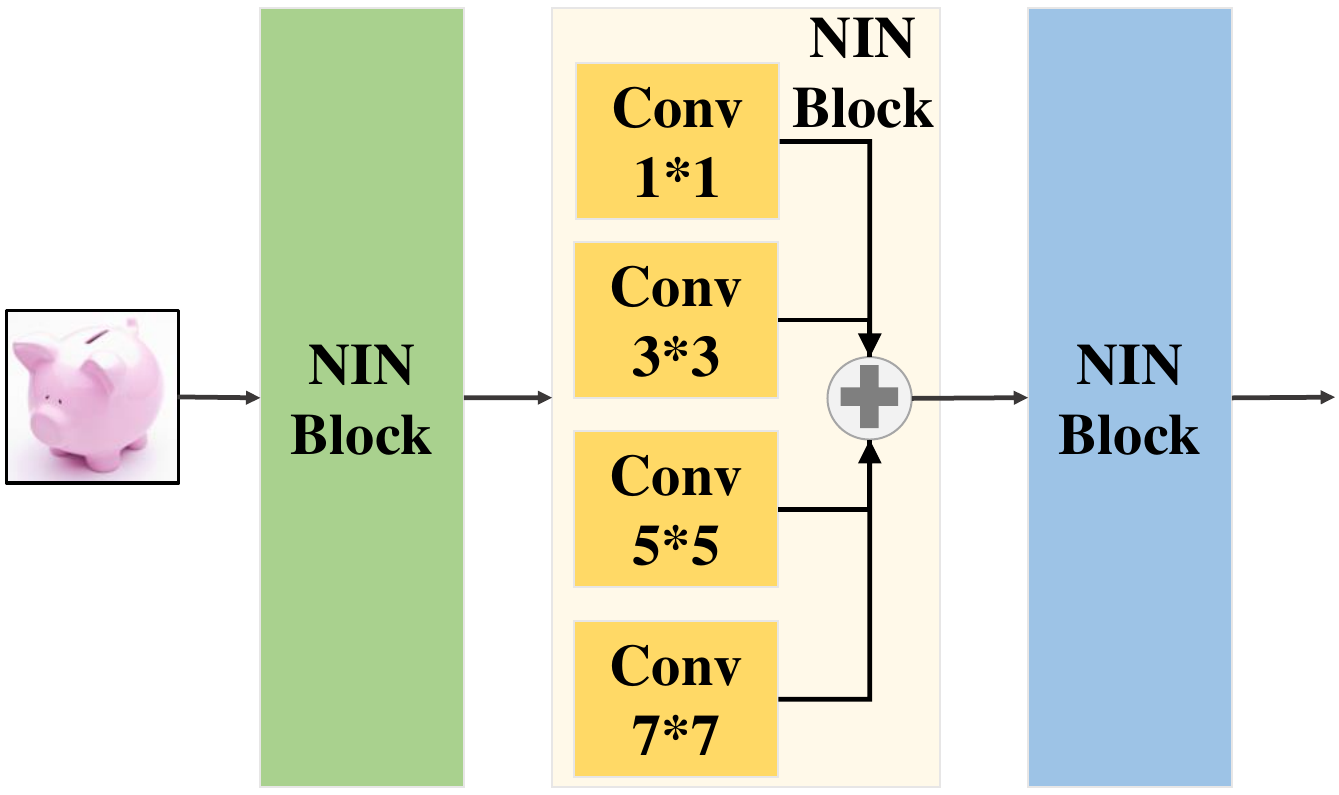}\caption{Net-in-Net}\label{pic:dense_connection}
\end{subfigure}
\vspace{-2mm}
\caption{Two Widely Used Model Structures.}
\label{pic:connection}
\vspace{-3mm}
\end{figure}

%Previous GIA works\cite{hatamizadeh2022gradvit,liu2022breaking,lu2022april} have evaluated self-attention\cite{vaswani2017attention}, encoding\cite{dosovitskiy2020image} structures in vision transformers, and found that the key to privacy leakage in vision transformer was the above structures involved. However, these conclusions were limited to vision transformers only, and there has been no discussion of the structures that are widely used by visual models in federated learning, such as skip connections\cite{he2016deep} (used such as ResNet, DenseNet families) and net-in-net\cite{szegedy2015going} (used such as GoogleNet and other InceptionNet families). Here, we start from widely used skip connections and explore how the model structure affects GIA.\par
\textit{\textbf{\Rmnum{1}. Skip connection}} is a widely used structure in deep neural networks that helps address gradient vanishing during training\cite{he2016deep}. It enables the flow of features from one layer to another by creating direct connections between non-adjacent layers. Generally, there are two common types of skip connections, derived from ResNets and DenseNets, as shown in Fig.~\ref{pic:connection}. In ResNet, skip connections take the form of identity mappings, where the input to a layer is \textbf{added directly} to the output of the subsequent layer. In contrast, DenseNet takes a more aggressive approach by densely \textbf{concatenating all previous layers within a block}, enhancing feature reuse \cite{huang2017densely}.\par

% \begin{figure}[ht]
%     %\vspace{-2mm}
%     \centering
%     \includegraphics[width=0.45\textwidth]{figs/backprob.pdf}
    
%     \caption{A Layer in Deep Neural Network with Three Types of Connections. 0) \textbf{no skip connection}, no additional operation; 1) \textbf{ResNet-like skip connection}, feature adding; 2) \textbf{DenseNet-like skip connection}, feature concatenation.}
%     \label{fig:backprob}
%     \vspace{-2mm}
% \end{figure}
To illustrate how skip connections affect the GIA, we start with a derivation on how they affect backpropagation. Assume a particular layer with $\mathbf{I}_{l}$ and $\mathbf{O}_{l}$ represent the input and output, respectively. We consider three types of connections: normal (sequential), ResNet-like, and DenseNet-like. And the process of backpropagation can be presented:\par
For \textbf{normal case}, $\mathbf{O}^{*}_{l}=\mathbf{O}_{l}$:\par
\begin{equation}
\label{eq:normback}
\frac{\partial\ell}{\partial \mathbf{I}_l}=\frac{\partial\ell}{\partial \mathbf{I}_{l+1}} \frac{\partial \mathbf{I}_{l+1}}{\partial \mathbf{I}_l}=\frac{\partial\ell}{\partial \mathbf{I}_{l+1}} \frac{\partial \mathbf{I}_{l+1}}{\partial \mathbf{O}_{l}} \mathbf{W}_{l}.
\end{equation}
For \textbf{ResNet} and \textbf{DenseNet}, the gradient in the deeper layers is passed to the front layers by adding or concatenation ( \scalebox{1.2}{$\boldsymbol{\oplus}$} represents both operations in Eq.~\eqref{eq:resback}):\par
\begin{equation}
\begin{split}
\label{eq:resback}
\frac{\partial\ell}{\partial \mathbf{I}_l} & = \frac{\partial\ell}{\partial \mathbf{I}_{l+1}} \frac{\partial \mathbf{I}_{l+1}}{\partial \mathbf{I}_l} = \frac{\partial\ell}{\partial \mathbf{I}_{l+1}} \frac{\partial \mathbf{I}_{l+1}}{\partial \mathbf{O}_l}\frac{\partial \mathbf{O}_l}{\partial \mathbf{I}_l} \\
& = \frac{\partial\ell}{\partial \mathbf{I}_{l+1}} \frac{\partial \mathbf{I}_{l+1}}{\partial \mathbf{O}_l}\frac{\partial (\mathbf{O}_l^{*}\oplus \mathbf{I}_l)}{\partial \mathbf{I}_l} = \frac{\partial\ell}{\partial \mathbf{I}_{l+1}} \frac{\partial \mathbf{I}_{l+1}}{\partial \mathbf{O}_l} (\mathbf{W}_{l} \underline{\oplus\textbf{1}}).
\end{split}
\end{equation}
Compared with Eq.~\eqref{eq:normback}, there is one more \textbf{residual term} in the gradient of Eq.~\eqref{eq:resback}. For models with skip connections, the residual terms multiply cumulatively during backpropagation, leading to a gradient that incorporates a greater number of combinations. This mechanism helps prevent gradient vanishing and \textbf{enhances the performance of GIAs by enabling them to utilize more information}. \par
\begin{figure}[t]
%\vspace{-2mm}
\centering
\begin{subfigure}{0.42\linewidth}\centering\includegraphics[width=1\linewidth]{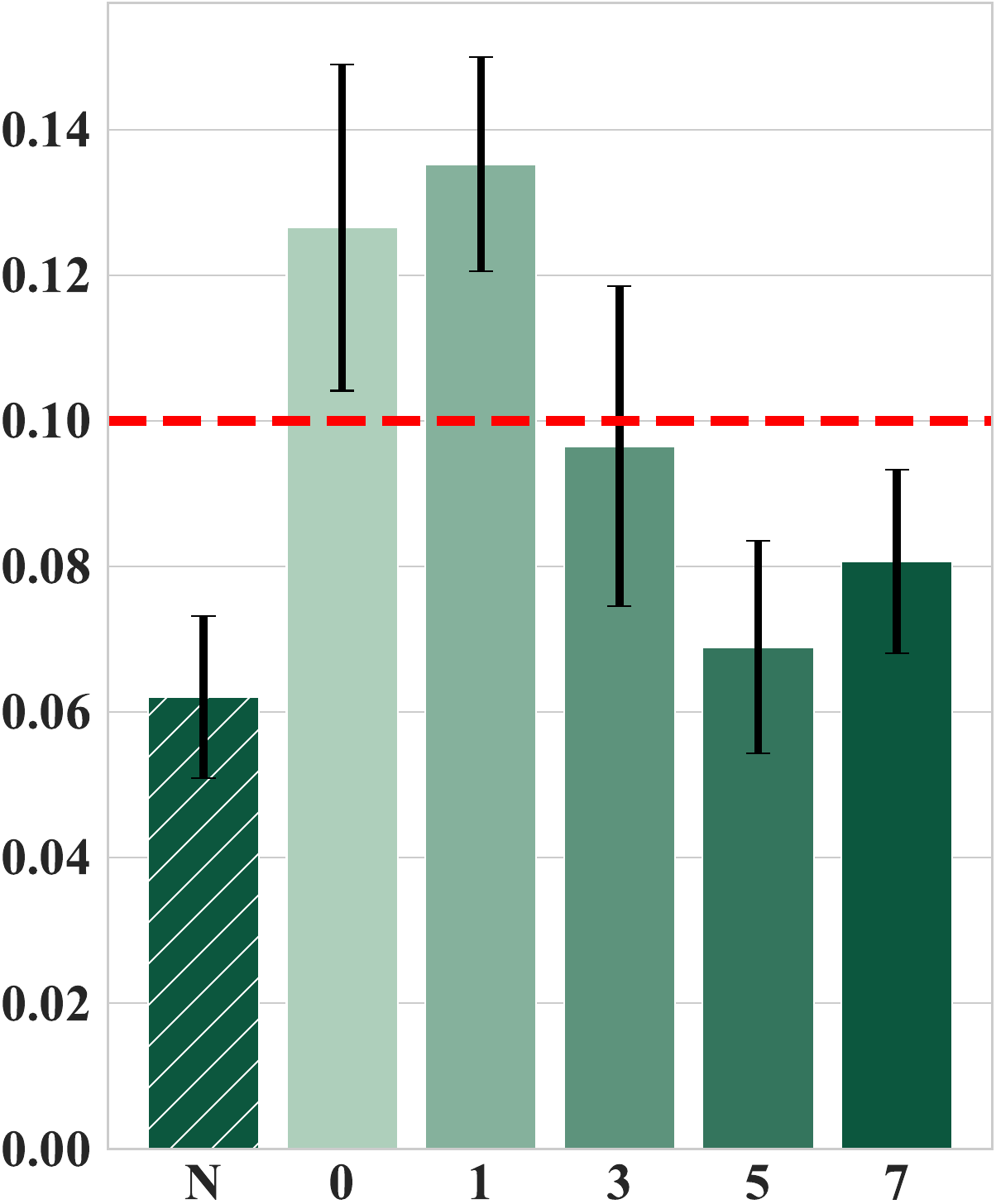}\caption{ResNet-18}\label{pic:skip-18}
\end{subfigure}
\centering
\begin{subfigure}{0.42\linewidth}\centering\includegraphics[width=1\linewidth]{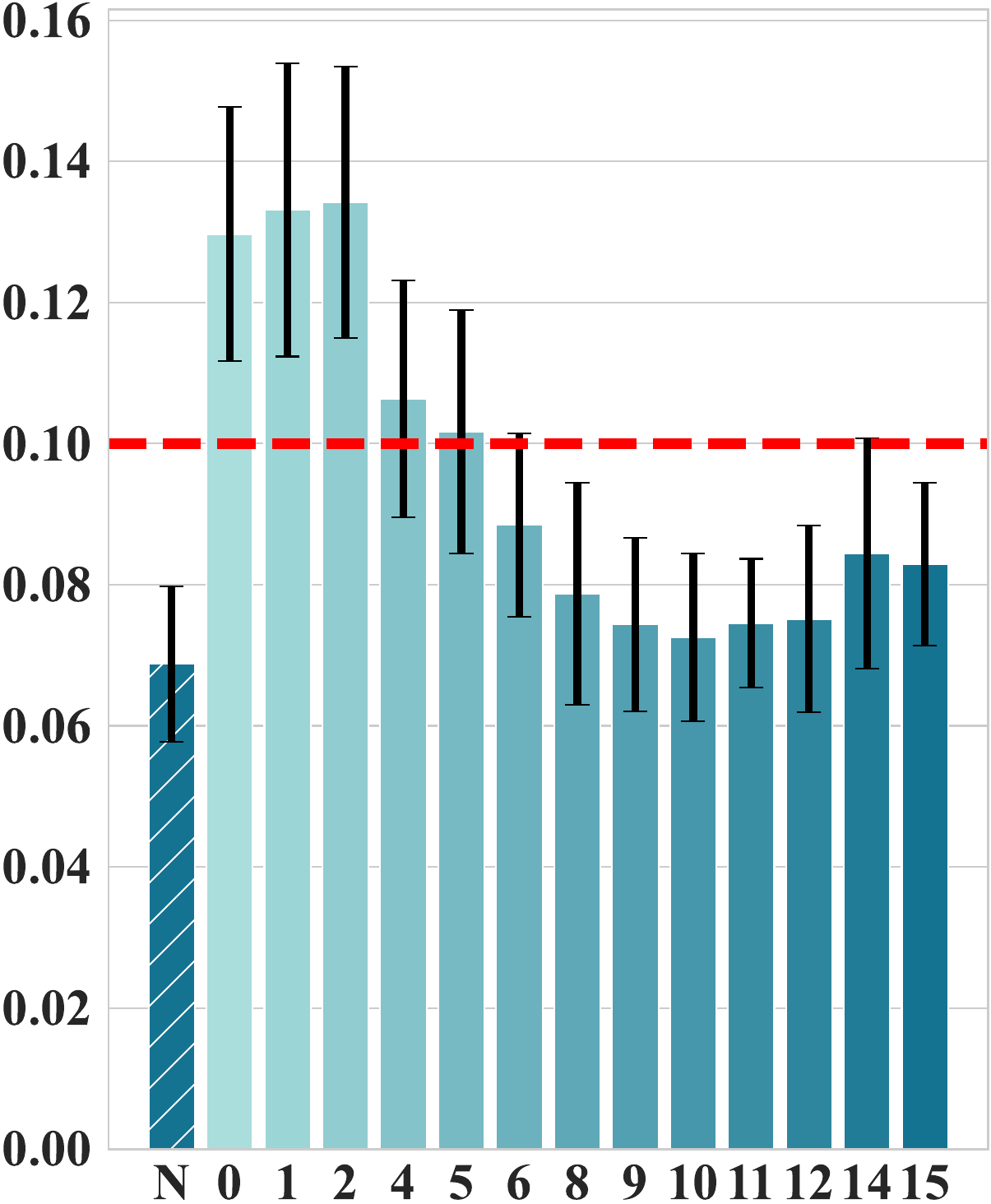}\caption{ResNet-34}\label{pic:skip-34}
\end{subfigure}
\vspace{-2mm}
\caption{Position of Skip Connections Affects GIA. \textbf{N} (\textbf{No Cut}) represents the original model, and thereafter \textbf{Id} represents the model cutting the \textbf{Id} connection, with darker colors representing deeper positions. \textbf{LPIPS}$\downarrow$.}
\label{pic:skip}
%\vspace{-2mm}
\end{figure}

We then verify the effect of skip connections on GIA based on ResNets and DenseNets, respectively. First, we evaluate how cutting skip connections at different positions affects the GIA on ResNet-18 and ResNet-34, as shown in Fig.~\ref{pic:skip}. We find that \textbf{(1) the presence of skip connections enhances the performance of the GIA}. Fig.~\ref{pic:skip} demonstrates that the original models are much more vulnerable to the GIA than others with connection cuts. Moreover, cutting the skip connection at any position significantly worsens the effectiveness of the GIA, even resulting in failed reconstruction (\textit{LPIPS} $ > 0.1$). In addition, we find that \textbf{(2) skip connections close to shallow layers have a greater impact on the GIA}. Fig.~\ref{pic:skip-34} shows that cutting shallow connections (e.g., $\#0,1,2,4,5$) drastically worsens the performance of the GIA. Recent works \cite{shwartz2017opening} regarding information flow in deep neural networks state that shallow layers are more sensitive to the input, and so are their gradients, which benefits GIAs.\par

\begin{figure}[t]
\vspace{-2mm}
    \centering
    \includegraphics[width=0.40\textwidth]{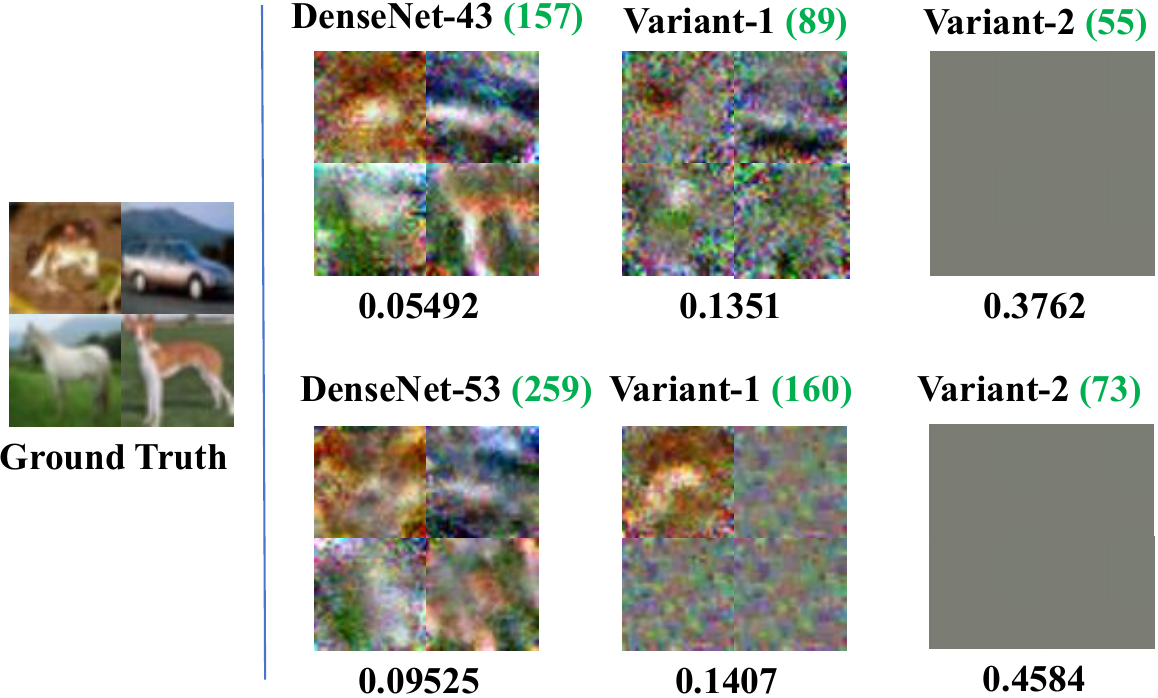}
    
    \caption{Number of Skip Connections Affects GIA. \textbf{Green number} represents the number of skip connections. \textbf{LPIPS}$\downarrow$.}
    \label{fig:densenet}
\vspace{-3mm}
\end{figure}

%their task is to perform feature extraction, while their gradients are designed to allow them to better understand the current input and therefore are more sensitive to the features.
Besides, we explore the impact of the number of skip connections on the GIA with DenseNets. We select DenseNet-43 and DenseNet-53 as baselines and obtain two variants for each by employing different cutting strategies\cite{ju2022connection}. As shown in Fig.~\ref{fig:densenet}, \textbf{reducing the number of skip connections greatly affects the performance of GIAs}. Images that can be easily inverted in \textit{baselines} are largely unrecognizable in \textit{variants-1}. Moreover, GIAs are completely unable to invert any information from \textit{variants-2}. Reducing the number of skip connections in a model decreases both the backpropagation paths and the residual terms. This leads to the gradients becoming less informative, which in turn limits the effectiveness of GIAs.\par
\begin{mdframed}[backgroundcolor=gray!10, roundcorner=5pt]
\textbf{(Insight \ref{sec:model-2-1})} (1) Skip connections alleviate gradient vanishing (\textbf{Pros}), while increasing the backpropagation paths and introducing residual terms, providing the adversary more information from gradients (\textbf{Cons}).
\end{mdframed}
%Based on the above analysis, the pros and cons of skip connections can be summarized as follows: on the one hand, skip connections alleviate gradient vanishing; but on the other hand, it increases the path of backpropagation and introduces residual terms, which enables the adversary to make use of more information from gradients to reconstruct input.\par
\textit{\textbf{\Rmnum{2}. Net-in-net}} (NIN), as proposed in \cite{szegedy2015going}, is a module that integrates multi-scale convolutional kernels within a single block, as illustrated in Fig.~\ref{pic:dense_connection}. In other words, NIN works like a widening layer, where more kernels capture richer input features. However, wider layers and multi-scale kernels also yield more informative gradients through backpropagation, consequently enhancing the effectiveness of GIAs.\par
% is proposed in \cite{szegedy2015going} as a module that incorporates multi-scale convolutional kernels within a single layer (as the right model in Fig.~\ref{fig:IandN}). NIN is essentially equivalent to a widening of layer, where the multi-scale convolutional kernel captures richer features from input. But it is clear that wider layers and multi-scale kernels also hold more informative gradients through backpropagation, thus benefiting GIA. 
% As Fig.~\ref{fig:IandN} shows, we perform GIA\cite{geiping2020inverting} on two models and find that replacing only the convolutional layers with NIN block of the same channels greatly improves the reconstruction quality. The adversary obtains more information from the gradient of the multi-scale convolutional layers, which makes the background, edges, and details of the reconstructed image clearer.\par
% \begin{figure}[ht]
%     \centering
%     \includegraphics[width=0.45\textwidth]{figs/IandN.pdf}
    
%     \caption{A Two-layer Model With/Without Net-In-Net Block And GIA's\cite{geiping2020inverting} Reconstructed Images with \textbf{LPIPS}$\downarrow$.}
%     \label{fig:IandN}
% \end{figure}
To further demonstrate the impact of NIN, we conduct ablation studies on GoogleNet and InceptionNet-V3, as detailed in Tab.~\ref{tab:inception}. Specifically, we compare the reconstruction results using the full gradients against those using only the gradients from a single NIN block. These findings validate that \textbf{the gradients from NIN blocks are crucial to the model's vulnerability to the GIA}. For example, as shown in Tab.~\ref{tab:inception}, although only the gradient of the first NIN block in GoogleNet, constituting merely  $2.27\%$ of the total parameters, was used, we unexpectedly achieved a result better than that obtained with the full gradient ($0.0879 < 0.2047$). This outcome occurs because approximating the full gradient is computationally intensive for an adversary and does not necessarily yield better reconstruction. In contrast, gradients of small, critical components, such as NIN blocks, are more likely to be approximated and leak inputs. \textbf{More broadly, auditing the risk of gradient leakage should focus more on identifying vulnerable structures within the gradients}. \par
\begin{table}[t]
\vspace{-2mm}
  \centering
  \renewcommand{\arraystretch}{0.75}
\setlength\tabcolsep{0.38pt}
  \caption{The Effectiveness of GIA Using Full or a Single NIN Block's gradient.}
  \footnotesize
  \setlength{\tabcolsep}{2pt} % 减小列之间的空白
  \begin{tabular}{ccccccc}
    \toprule
    \multirow{1.5}[4]{*}{\textbf{Model}} & \multirow{1.5}[4]{*}{\textbf{Metric}} & \multicolumn{5}{c}{\textbf{The ID of NIN Blocks Providing Gradient}} \\
    \cmidrule{3-7}
    %\cmidrule{1}
    & & Full & \#1 & \#2 & \#3 & \#4 \\
    \midrule
    \multirow{0.75}[4]{*}{GoogleNet} & \textit{Params Ratio} & 100.00\% & 2.72\% & 11.68\% & 6.58\% & 9.25\% \\
    & \textit{LPIPS$\downarrow$} & 0.2047 & \textbf{0.0879} & \textbf{0.0776} & \textbf{0.0859} & \textbf{0.0756} \\
    \midrule
    \multirow{0.75}[4]{*}{Inception V3} & \textit{Params Ratio} & 100.00\% & 1.14\% & 1.56\% & 1.76\% & 5.26\% \\
    & \textit{LPIPS$\downarrow$} & 0.2672 & \textbf{0.0858} & \textbf{0.0897} & \textbf{0.0875} & \textbf{0.0804} \\
    \bottomrule
  \end{tabular}%
  \label{tab:inception}%
  %\vspace{-3mm}
\end{table}%
%The pros and cons of NIN are summarized as follows: on the one hand, NIN uses multi-scale convolution for better feature extraction, but on the other hand, it is precisely this backpropagation that requires different updates to multiple convolutional kernels, which makes the gradient contain more information and is more prone to privacy leakage.
\begin{mdframed}[backgroundcolor=gray!10, roundcorner=5pt]
\textbf{(Insight \ref{sec:model-2-1})} (2) NIN utilizes multi-scale kernels for stronger feature extraction capability (\textbf{Pros}), while requiring different updates to multiple kernels, which provides informative gradients to benefit GIAs (\textbf{Cons}).
\end{mdframed}

\subsubsection{Micro Design: Tiny Clue Reveals General Trend}
\label{sec:model-2-2}
Micro designs are subtle techniques ubiquitous in nearly all modern models. To investigate their impact on the GIA, we evaluate six prevalent micro designs: \textit{bias}, \textit{activation function (ReLU)}, \textit{dropout}, \textit{max pooling}, \textit{convolutional kernels (size)}, and \textit{padding}. In particular, we make a series of modifications to a configurable model, ConvNet\cite{geiping2020inverting}, which only includes components related to micro designs ensuring a clean setting. The standard ConvNet incorporates bias, employs ReLU functions, and includes two max pooling layers and one dropout layer. Additionally, all convolutional layers are equipped with $3\times3$ kernels and padding $1$. Details are provided in Appx.~\ref{appx:sup}.\par
 
Our findings indicate that \textbf{micro designs significantly impact the model's resistance to the GIA}. As shown in Tab.~\ref{tab:modifications}, removing \textit{ReLU}, dropout, \textit{max pooling} layers, or increasing the \textit{kernel size} substantially exacerbates the model's vulnerability to GIAs. In contrast, removing \textit{bias}, decreasing the \textit{kernel size}, or expanding \textit{padding} enhances the resilience.\par
\begin{table}[t]
  \vspace{-2mm}
  \centering
    \renewcommand{\arraystretch}{0.3}
\setlength\tabcolsep{0.38pt}
  \caption{The Effectiveness of GIA against Model Modifications. (+): Modifications improve the reconstruction quality, while others diminish it (--). (\textbf{R})emove, (\textbf{I})ncrease, (\textbf{D})ecrease.}
  \footnotesize
    \begin{tabular}{ccccccc}
    \toprule
    \textbf{Mods} & None  & \textbf{R} ReLU & \textbf{R} DropOut & \textbf{R} MaxPool2d & \textbf{I} kernel to $4$ \\
    \midrule
    \textbf{LPIPS$\downarrow$} & 0.0044(+) & 0.0011(+) & 0.0043(+) & 0.0000(+) & 0.0012(+)\\
    \midrule
    \textbf{Mods} & \textbf{R} bias  & \textbf{D} kernel to $2$ & \textbf{D} kernel to $1$ & \textbf{I} padding to 2 & \textbf{I} padding to 3 \\
    \midrule
    \textbf{LPIPS$\downarrow$} & 0.0327(--) & 0.0299(--) & 0.1507(--) & 0.0071(--) & 0.0506(--) \\
    \bottomrule
    \end{tabular}%
  \label{tab:modifications}%
  %\vspace{-3mm}
\end{table}%
Essentially, micro design affects the amount of information available in the feature map. Specifically, \textbf{reducing the information related to the input in feature maps would render GIAs less effective}. (1) Feature map sparsification. The \textit{activation function} and \textit{padding} transform or zero out elements in the feature map. (2) Feature map aggregation. The \textit{max pooling} layer selects representative elements, and a smaller \textit{kernel size} focuses on more localized features. These designs reduce the correlation between the input and the feature maps, and thus affect the accurate inversion of input data from gradients.
However, \textbf{enhancing the available information contained in the feature map benefits GIAs}. Increasing \textit{kernel size} would promote the model in extracting features on a broader scale, thereby containing more information, while \textit{bias} provides extra parameters for the adversary.
\begin{mdframed}[backgroundcolor=gray!10, roundcorner=5pt]
\textbf{(Insight \ref{sec:model-2-2})} Micro designs influence the available information about the input in feature maps, consequently affecting the gradient and the effectiveness of GIAs.
%The micro designs affect how much the available information is between the model feature maps and the input, which in turn affects the gradient and has a significant impact on the GIA.
\end{mdframed}

\section{Evaluation on Post-Processing}
\label{sec:post-processing}
% In practical FL systems, clients often apply post-processing techniques to gradients before sharing them with the server.  These methods serve to obfuscate the shared gradients and potentially provide defense for clients against GIAs. This section examines the effectiveness of four commonly utilized post-processing techniques in defending against GIAs under a practical FL setting. Moreover, we evaluate their capacity to address the critical trade-off between the model utility and defensive performance.
In practical FL systems, clients often apply post-processing techniques to gradients before sharing them. These methods obfuscate the shared gradients to offer potential defense for clients against GIAs. In this section, we study the effectiveness of four commonly utilized post-processing techniques in defending against GIAs under a practical FL setting. Moreover, we evaluate their capacity to address the critical trade-off between the model utility and defensive performance. \par
\textbf{Experimental setup:} We consider a practical FL system that involves $100$ clients collaboratively training ResNet-18 and Swin~\cite{liu2021swin} models on CIFAR10 and CIFAR100 datasets. The server launches three SOTA GIAs: \textbf{GIA-O}, \textbf{GIA-L}, and \textbf{ROG}\cite{yue2023gradient}, a recent GIA that breaks through gradient obfuscation. To defend against these attacks, the client employs four post-processing techniques on the gradient: quantization \textbf{(Q)}\cite{alistarh2017qsgd}, sparsification \textbf{(S)}\cite{eghlidi2020sparse}, clipping \textbf{(C)}\cite{mcmahan2017learning}, and perturbation \textbf{(P)}\cite{naseri2020local}. Details are provided in Appx.~\ref{appx:setup}. \par

\begin{figure}[t]
%\vspace{-2mm}
    \centering
    \includegraphics[width=0.42\textwidth]{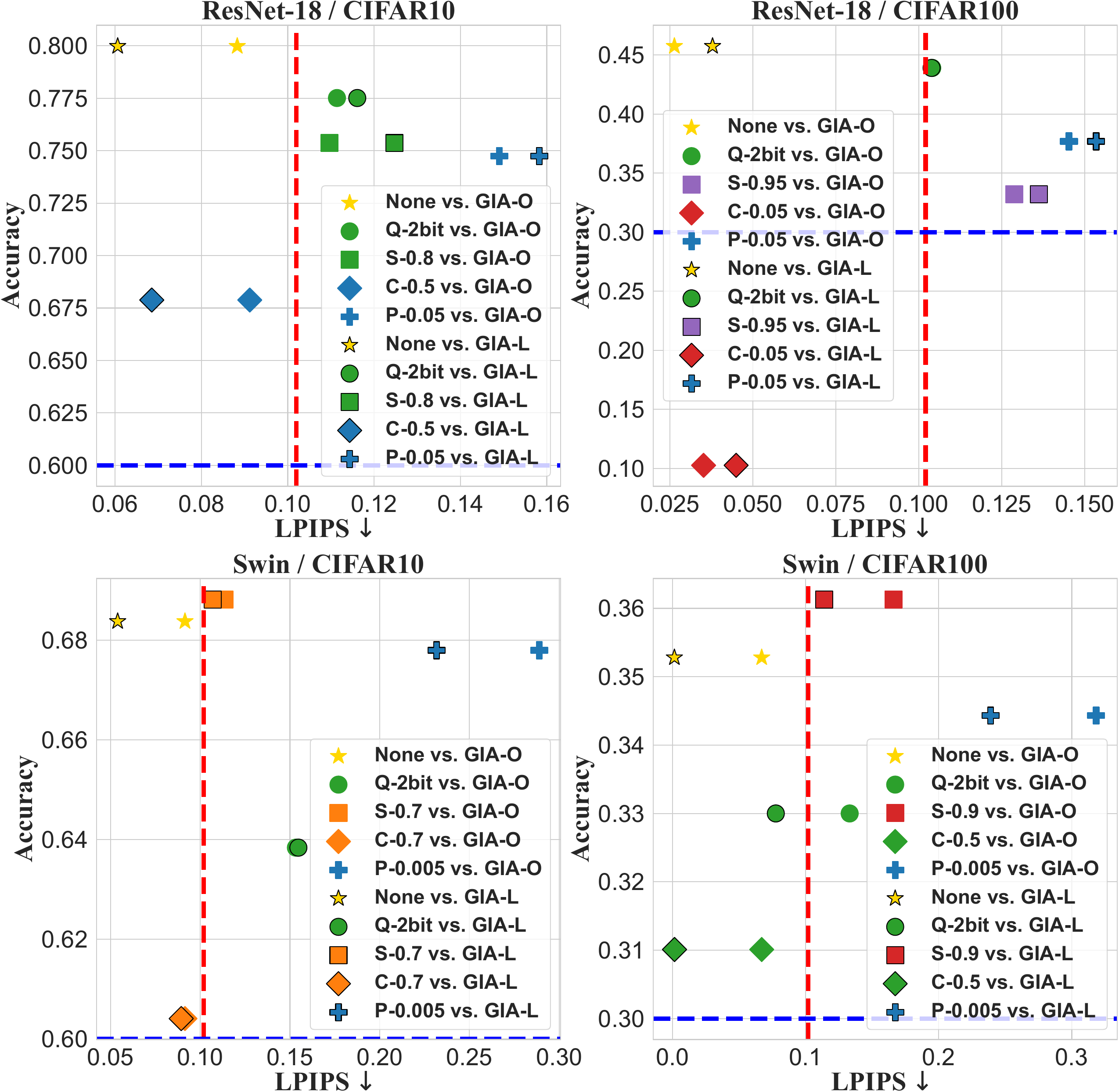}
    
    \caption{Best Results That Maintain Privacy-Utility Tradeoff in FL Systems Utilizing Various Post-Processings. \textbf{Red Line} is a split line of privacy leakage. \textbf{Blue Line} is a split line of model's acceptable utility.}
    \label{fig:tradeoff}
\vspace{-2.0mm}
\end{figure}

We conduct experiments to assess the effectiveness of four post-processing techniques against GIAs and their impact on accuracy under different parameter settings. We choose the optimal performance for each post-processing technique, representing the best privacy-utility trade-off in Fig.~\ref{fig:tradeoff}. The results of ROG are presented in Tab.~\ref{tab:rog}, Appx.~\ref{appx:sup}. We show that: \textbf{most post-processing techniques can effectively defend against the strongest GIAs without significantly compromising accuracy}. This is illustrated by their points distributed above the blue line (accuracy degradation of no more than $30\%$) and to the right of the red line. Among these techniques, quantization demonstrates the most favorable trade-off. In the experiment involving ResNet-18 and CIFAR100 (Fig.~\ref{fig:tradeoff}), utilizing $2$-bit quantization on shared gradients enables clients to defend against GIAs with an accuracy drop of no more than 5\%. However, clipping fails to guarantee the trade-off. As shown in Fig.~\ref{fig:tradeoff}, even with a large clip value that significantly impacts accuracy, it has almost no effect on GIAs. Overall, the post-processing methods demonstrate effective defense capabilities against GIAs while avoiding the significant performance degradation or computational and communication overheads that may be associated with traditional defense mechanisms such as secure multi-party computation (discussed in the Appx.~\ref{appx:sup}).

%\begin{figure}[ht]
%    \centering
%    \includegraphics[width=0.45\textwidth]{figs/summary_plot.pdf}
%    \caption{Privacy-Utility Tradeoffs in FL Systems Using Various Post-Processings}
%    \label{fig:tradeoff}
%\end{figure}
\begin{mdframed}[backgroundcolor=gray!10, roundcorner=5pt]
\textbf{(Insight \ref{sec:post-processing})} In practical FL systems, even trivial post-processings applied to gradients can easily defend against the most powerful GIAs while maintaining model accuracy.
\end{mdframed}

\section{Discussion}
\label{sec:dis}
In this section, we summarize our systematization and evaluation, and provide recommendations for the future exploration and mitigation of gradient leakage risks in FL systems.

\textbf{Explore the risks of GIAs in practical FL systems.} Although previous research has demonstrated that GIAs pose significant threats, these notable effects rely on assumptions that the attacks are either overpowered or not fully integrated with the FL system. Our evaluations indicate that GIAs are limited and fragile in practice, suggesting several promising directions for future research: \textbf{(1) Investigating the update leakage beyond the gradient.} Future studies should focus on more practical local training settings (e.g., mini-batch updates), exploring the reconstruction of local data based on parameter updates instead of raw gradients. \textbf{(2) Assessing GIA risks across practical scenarios.} Most existing GIAs are focused on image classification tasks. However, the risks associated with other practical tasks—such as federated instruction fine-tuning~\cite{zhang2024towards}—as well as diverse modalities and advanced architectures (e.g., large language and vision models) remain underexplored. \textbf{(3) Developing robust GIAs.} In real-world applications, shared gradients are often post-processed or defended. Future research could explore reconstructing local data with inaccurate gradients.

\textbf{Mitigate the risks of GIA through inherent and lightweight mechanisms.} By evaluating three key aspects, we demonstrate that GIA's effectiveness can be easily compromised. Consequently, we advocate for defenses against GIA that leverage the inherent properties of the FL system or utilize lightweight mechanisms, rather than utilizing subtle and complex designs. \textbf{(1) Exploring the natural defensive potential of practical FL systems.} Future applications could focus on designing FL systems that are inherently resistant to GIAs. For example, secure local training settings can be configured by increasing the number of local update rounds or augmenting the client training data, such as by injecting public samples. \textbf{(2) Developing fine-grained defense mechanisms.} To prevent over-defense from compromising utility, future research could identify vulnerable phases within the FL training process or pinpoint key structures (e.g. skip connections) within the model for targeted defense. \textbf{(3) Enhancing post-processing techniques.} Given that even basic post-processing techniques demonstrate substantial potential in defending against GIAs, future efforts could focus on further improving their effectiveness in balancing the privacy-utility tradeoffs, e.g., designing adaptive post-processing methods.

\textbf{Gradient Leakage on language models.}  While existing GIAs primarily focus on image classification tasks, investigating the reconstruction of textual data on language models is crucial for advancing future research, particularly in the era of large language models (LLMs) and multimodal systems. In this context, we provide some concerns and outlooks on this topic: \textbf{(1) Metrics for privacy leakage in textual data:} Unlike image data, textual data lacks robust metrics to quantify privacy leakage for reconstruction attacks. For a reconstructed text, ``look-alike'' does not necessarily indicate privacy leakage if the semantic meaning or critical information is altered. For example, reconstructing ``The cat is on the mat'' as ``The cat is on the hat'' introduces minimal character changes but significantly alters the meaning, leading to a little privacy leakage. Therefore, developing fair metrics is essential for auditing GIAs on textual data. \textbf{(2) Emphasis on text generation tasks: }Existing GIAs tend to focus on text classification tasks~\cite{deng2021tag,gupta2022recovering,balunovic2022lamp}, yet language models in practical FL systems are mainly trained for generation tasks (e.g., Q\&A). These tasks differ fundamentally in gradient computation: while classification tasks calculate gradients after processing the entire input, generation tasks compute and accumulate gradients word-by-word in an auto-regressive manner. This distinction introduces unique challenges for applying GIAs to text generation tasks. \textbf{(3) Challenges with large language models (LLMs):} Recent studies have highlighted that LLMs, pre-trained or fine-tuned in FL settings, are potentially vulnerable to gradient leakage~\cite{ye2024openfedllm}. However, implementing GIAs on LLMs presents significant obstacles. First, LLMs are trained with large-batch, high-dimensional datasets—for instance, GPT-3-175B training involves batch sizes of up to 3.2 million tokens~\cite{brown2020language}—which surpass current GIA capabilities. Second, the complexity of LLM architectures and their vast parameter counts complicate gradient inversion. Finally, communication constraints often result in clients uploading only partial gradients during training, frequently employing proxy models or parameter-efficient fine-tuning (PEFT) techniques~\cite{fan2023fate}. This severely restricts the information available to adversaries.

\section{Conclusion}
In this work, we conducted a comprehensive study on GIAs in practical FL systems. We thoroughly reviewed the evolution of the GIA, highlighting the key milestones and breakthroughs. Additionally, we established a systematization of GIAs to reveal their inherent threats. We indicated that the notable effectiveness demonstrated by current GIAs relies on ideal settings with auxiliary assumptions. To evaluate the actual threat of GIAs against practical FL systems, we identified three fundamental aspects influencing GIAs's effectiveness: \textit{training setup}, \textit{model}, and \textit{post-processing}. Through theoretical and empirical evaluations of SOTA GIAs in diverse settings, our findings indicate that GIAs are \textit{constrained}, \textit{fragile}, and \textit{easily defensive} in practice. The actual threats posed by GIAs to practical FL systems are limited, despite their perceived potency in previous literature. We hope our work corrects some misconceptions and promotes more precise and realistic investigations into GIAs within FL systems.\par
\section*{Acknowledgments}
This work is supported by National Natural Science Foundation of China (Grant No. U24B20182, 62122066, 62472158, 62102337),  National Key R\&D Program of China (Grant No. 2021ZD0112803), Key R\&D Program of Zhejiang (Grant No. 2024C01164, 2022C01018), Young Elite Scientists Sponsorship Program by CAST (Grant No. 2023QNRC001), and Natural Science Foundation of
Hunan Province, China (Grant No. 2023JJ40174).
\section*{Ethical Considerations}
The authors have reviewed the ethical considerations outlined in the conference documents, including the \textit{Call for Papers}, \textit{Submission Policies and Instructions}, and \textit{Ethics Guidelines}. We confirm that all datasets and models utilized in our experiments are open-source, publicly available, and non-sensitive. Our evaluations did not involve the disclosure of harmful or sensitive content to the public or other researchers. Moreover, our study did not introduce any new risks; instead, it highlighted that existing GIAs are constrained, fragile, and easily defensible in FL systems, thereby contributing to a better understanding of gradient leakage.
\section*{Compliance with the Open Science Policy}
In full compliance with the Open Science Policy, we commit to sharing all research artifacts associated with this study, including datasets, scripts, and source code. These resources are publicly available via a Zenodo repository: https://zenodo.org/records/14664682.

\bibliographystyle{plain}
\bibliography{sample}
%\newpage
\appendices
\section{Proofs}
\label{appx:proof}
\textbf{\Rmnum{1}. Proof for Lemma.~\ref{lemma:gdx}:} 
Based on Eq.~\eqref{eq:mu}, \eqref{eq:f_l} and \eqref{eq:loss},  we can derive the gradients of every layer for the following iterative form:
\begin{equation}
\label{eq:gradients}
\begin{aligned}
    \frac{\partial\ell}{\partial \mathbf{W}_{L}} &= y\frac{\partial\ell}{\partial\mu}\mathcal{F}_{L-1}^\top, \\
    \frac{\partial\ell}{\partial \mathbf{W}_{L-1}} &= \left((\mathbf{W}_{L}^\top(y\frac{\partial\ell}{\partial\mu}))\odot\sigma_{L-1}^{\prime}\right)\mathcal{F}_{L-2}^\top, \\
    \frac{\partial\ell}{\partial \mathbf{W}_{L-2}} &= 
    \left(\mathbf{W}_{L-1}^\top\left(\left(\mathbf{W}_L^\top\left(y\frac{\partial\ell}{\partial\mu}\right)\right)\odot\sigma_{L-1}^{\prime}\right)\right) \\
    & \quad \odot\sigma_{L-2}^{\prime}\mathbf{P}^\top.
\end{aligned}
\end{equation}
Particularly, from Eq.~\eqref{eq:mu} and \eqref{eq:gradients}, we can characterize the dependence between the gradient of the last layer $\frac{\partial\ell}{\partial \mathbf{W}_{L}}$ and $\mu$ as follows\cite{zhu2020r}:
\begin{equation}
\label{eq:depend}
    \begin{aligned}
\frac{\partial\ell}{\partial \mathbf{W}_L}\mathbf{W}_L=y\frac{\partial\ell}{\partial\mu}\mathcal{F}_{L-1}^\top\mathbf{W}_L=\frac{\partial\ell}{\partial\mu}\mu=\frac{-\mu}{1+e^\mu}.
\end{aligned}
\end{equation}
With $\mu$, the input $\mathbf{x}$ can be determined by iteratively solving Eq.~\eqref{eq:mu}, as we derived in Eq.~\eqref{eq:mux}.\par
\noindent\textbf{\Rmnum{2}. Proof for Theorem.~\ref{the:updates}:} Based on Eq.~\eqref{eq:error}, the error between the ground truth $\mathbf{x}^*$ and the reconstructed data $\mathbf{x}^{rec}$ is determined by the outputs:
\begin{equation}
\label{eq:gtmu}
    \begin{aligned}
\mathcal{G}\left(\mu^*\right)=\frac{-\mu^*}{1+e^{\mu^*}}=\frac{\partial\ell}{\partial \mathbf{W}^0_L}\mathbf{W}^0_L,
\end{aligned}
\end{equation}
\begin{equation}
\label{eq:recmu}
    \begin{aligned}
\mathcal{G}\left(\mu^{rec}\right)=\frac{-\mu^{rec}}{1+e^{\mu^{rec}}}=\left(\frac{\partial\ell}{\partial \mathbf{W}_L^0}+\sum_{u=1}^{U-1}\frac{\partial\ell}{\partial \mathbf{W}_L^u}\right)\mathbf{W}^0_L.
\end{aligned}
\end{equation}
The output difference can be approximated by:
\begin{equation}
\label{eq:recmingt}
\begin{aligned}
\mu^{rec}-\mu^{*}\approx\frac{\mathcal{G}\left(\mu^{rec}\right)-\mathcal{G}\left(\mu^{*}\right)}{\mathcal{G}'\left(\mu^{*}\right)}=\frac{(e^{{\mu^{*}}}+1)^{2}\Sigma_{u=1}^{U-1}\frac{\partial\ell}{\partial \mathbf{W}_{L}^{u}}\mathbf{W}^0_{L}}{\mu^{*}e^{{\mu^{*}}}-e^{{\mu^{*}}}-1}.
\end{aligned}
\end{equation}
\par
\noindent\textbf{\Rmnum{3}. Proof for Theorem.~\ref{the:bound}:} By the Lagrange Mean Value Theorem, for a given mask $\Delta$, we can find a $\xi$ such that: 
\begin{equation}
\label{eq:lagrange}
\begin{aligned}
\Phi\left(\mathbf{x}+\Delta\right)=\Phi\left(\mathbf{x}\right)+\Phi'\left(\mathbf{\xi}\right)\Delta.
\end{aligned}
\end{equation}
Assume that $\Delta$ and $\mathbf{x}$ are linearly related and that $\mathbf{x}+\Delta$ is neighboring $\mathbf{x}$:
\begin{equation}
\label{eq:applagrange}
\begin{aligned}
\Phi\left(\mathbf{x}+\Delta\right)-\Phi\left(\mathbf{x}\right)=\Phi'\left(\mathbf{\xi}\right)\epsilon\mathbf{x}\approx\Phi'\left(\mathbf{x}\right)\epsilon\mathbf{x},
\end{aligned}
\end{equation}
where $\epsilon$ is the scale factor.
$\Phi\left(\mathbf{x}+\Delta\right)-\Phi\left(\mathbf{x}\right)=\mathbf{0}$ if and only $\Phi'\left(\mathbf{x}\right)\mathbf{x}=\mathbf{0}$, which indicates $\Phi'\left(\mathbf{x}\right)$ is not full rank. Considering  $\Phi'\left(\cdot\right) \in \mathbb{R}^{p \times B \times D}$, it follows that there must exist at least one nonzero solution when $p < B\times D$.

\section{Experimental Setups}
\label{appx:setup}
\textbf{\Rmnum{1}. Details for experiments in Sec.~\ref{sec:training}:} 
We conduct evaluations on CIFAR10 (at resolutions of $32\times32$ and $64\times64$) and ImageNet-1K (at resolutions of $128\times128$, $256\times256$, and $512\times512$) datasets, employing $B=1$ for Fig.~\ref{pic:pixel}. Subsequently, for Fig.~\ref{pic:bs}, we assess GIAs using the CIFAR100 dataset with a fixed resolution of $32\times32$. In GIA-O, the ratios for priors are specified as $1e^{-5}$. For GIA-L, we adhere to Jeon's approach\cite{jeon2021gradient} involving two-stage optimization, comprising DcGAN pretraining for CIFAR10/100 and BigGAN for ImageNet-1K. We opt for the Adam optimizer (8000 iterations for GIA-O, 400 iterations for latent space optimization and 6000 iterations for optimizing the generator in GIA-L). The evaluation of GIAs is performed on the ResNet-18 model\cite{he2016deep} initialized with PyTorch's default initialization.\par
%We perform evaluations on CIFAR10\cite{krizhevsky2009learning} ( $32\times32,64\times64$) and ImageNet-1K\cite{deng2009imagenet} ($128\times128,256\times256,512\times512$) datasets with $B=1$ for Fig.~\ref{pic:pixel}. And for Fig.~\ref{pic:bs}, we evaluate GIA on CIFAR100 dataset\cite{krizhevsky2009learning} with fixed resolution $32\times32$. For GIA-O, the ratios for priors are set to $1e^{-5}$.  For GIA-L, we follow Jeon's\cite{jeon2021gradient} setting with two-stage optimization (pretraining DcGAN\cite{radford2015unsupervised} for CIFAR10/100 and BigGAN\cite{brock2018large} for ImageNet-1K). We utilize Adam optimizer\cite{kingma2014adam} for optimization (8000 iterations for GIA-O and 400 iterations for GIA-L's latent space optimization, 6000 for optimizing generator). GIAs are evaluated on ResNet-18 model\cite{he2016deep} with PyTorch's default initialization (Xavier\cite{glorot2010understanding}). For Fig.~\ref{fig:OOD}, GIA-L utilizes 4000 iterations of latent space optimization in order to maximize the expressive capability of the generator.\par%, together with learning rate scheduling\cite{goyal2017accurate} and early stopping\cite{prechelt2002early} mechanisms to get the best result.\par
%Besides, we use \textit{Learned Perceptual Image Patch Similarity (LPIPS)}\cite{zhang2018unreasonable}, implementing by pre-trained Vgg-11 model\cite{simonyan2014very}, as the attack evaluation metric.
\noindent\textbf{\Rmnum{2}. Details for experiments in Sec.~\ref{sec:model}:} We adopt Geiping's GIA\cite{geiping2020inverting} as the baseline, incorporating solely gradient matching and $p_{TV}$ as loss function. Nine models are trained on the CIFAR10 dataset with a batch size of $128$ and a learning rate of $0.01$ for $10$ local epochs. The whole FL process involves $5$ clients for $100$ rounds totally. The For each model, we calculated its IGSA values by averaging the reconstruction results of $10$ independent samples.  with parameters $K=10000$ and $r=1e^{-3}$, and subsequently normalized for each model.\par
%To focus on model, we use Geiping's GIA\cite{geiping2020inverting} as baseline with only gradient matching and $p_{TV}$ as loss. We simulate a FedAvg setting and train 100 rounds (learning rate $0.1$) on the CIFAR10 dataset for 9 models with 6 clients, consisting of five normal client ($N=10000$, $B=64$, $E=10$) and one victim client ($N=B=1$, $E=1$) for evaluating GIA. Each IGSA value is averaged with 100 samples in CIFAR10 with $K=10000$ and $r=1e^{-3}$. IGSA values for each model are normalized.\par
\noindent\textbf{\Rmnum{3}. Details for experiments in Sec.~\ref{sec:post-processing}:} We adopt a practical FedAvg setup where, in each round, $10\%$ of clients are randomly chosen for training, with $E=1$ and $B=10$. To optimize GIA's performance, our client pool comprises 99 normal clients ($N=500$) and 1 victim client ($N=10$).
%We consider a practical FedAvg setting. In each round, $10\%$ clients are randomly selected for training with local settings, $E=1$ and $B=10$. To maximize GIA's performance, the clients consist of 99 normal clients ($N=500$) and 1 victim client ($N=10$). GIA-O utilizes both $p_{BN}$ and $p_{TV}$. Other settings are same as Sec.~\ref{sec:training}.\par

\section{Supplementary Materials}
\label{appx:sup}
%\textbf{\Rmnum{1}. Introduction to micro designs: } \textbf{(1) Bias} terms are added to each neuron in layers to allow the model to capture patterns that might not be centered around zero. They help shift the activation function and make the network more expressive\cite{lecun1998gradient}. \textbf{(2) Activation function} introduces non-linearity to neural network. Among the commonly used activations, Rectified Linear Unit (ReLU)\cite{glorot2011deep} is widely used for its simplicity and effectiveness in training deep networks. \textbf{(3) Dropout} is a regularization technique that randomly sets a fraction of neurons' outputs to zero during training\cite{srivastava2014dropout}. This prevents overfitting and encourages the network to learn more robust features. \textbf{(4) Max pooling} is a down-sampling technique used in convolutional neural network (CNN). It reduces the spatial dimensions of feature maps by selecting the maximum value from a group of adjacent values, which helps the network focus on the most important features. \textbf{(5) Convolutional kernels} are small filters that slide over input data to extract features\cite{lecun1998gradient}. In CNNs, these kernels are used to perform convolutions, capturing spatial hierarchies and patterns in the data. \textbf{(6) Padding} is the process of adding extra elements (usually zeros) around the input data before convolution or pooling operations. It helps maintain the spatial dimensions of the input and prevents the reduction in feature map size.\par
\textbf{\Rmnum{1}. Related works:} The most closely related works can be categorized into two main groups: (1) Survey on GIAs. Prior surveys\cite{yang2023gradient,zhang2022survey} primarily focused on a coarse-grained categorization of GIAs with optimization-based and analysis-based, providing their definitions. In this paper, we go a step further by summarizing the evolution of GIAs, marking a key milestone, and offering a detailed systematization. Specifically, we characterize the threat models, categorizing the adversary assumptions so far. (2) Evaluation on GIAs. Previous works\cite{huang2021evaluating,hatamizadeh2023gradient} have highlighted that the lack of a strong assumption (BN statistic) affects the efficacy of GIAs in practice. However, they have not systematically identified the key challenges that practical FL systems pose to GIAs and given a comprehensive analysis. Yue et al.\cite{yue2023gradient} have pointed out that techniques such as clipping and compression are ineffective in defending GIAs, which appears to contradict the conclusions in Sec~\ref{sec:post-processing}. This discrepancy arises because they did not conduct a thorough evaluation of the defensive performance of these methods in a practical FL setting. Besides, Wang et al. ~\cite{wang2024more} indicate that the effectiveness of GIA in practice is limited. However, their evaluations are restricted to the impact of model initialization and the number of local updates on two earlier GIA-Os with comparatively weaker effectiveness. In contrast, we evaluate several state-of-the-art GIAs from the GIA-O and GIA-L categories, and present extensive evaluations from various aspects, uncovering findings into the limited threats posed by GIAs in practice comprehensively.\par
% The reason is that our experimental setups are different. In this work, we aim to investigate the "privacy-utility'' trade-off of the four post-processing techniques in practical FL training. While Yue et al, on the other hand, only proved that the clipping technique does not guarantee the trade-off (which is consistent with one of our conclusions in Sec.~\ref{sec:post-processing}). And they did not give experimental results for the other three techniques in the practical setting.\par

\noindent\textbf{\Rmnum{2}. Discussion on metric selection:} In this paper, we primarily use the averaged $LPIPS$ metric to evaluate the privacy leakage of the batch of reconstructed images, which is widely adopted in previous works. \textit{While Carlini et al.~\cite{carlini2022membership} argue that averaged metrics are inadequate for membership inference attacks due to the potential presence of partially vulnerable samples, the averaged metric remains appropriate for GIAs. This is because reconstructed samples within a batch typically exhibit similar reconstruction quality}. 

To support this, we introduce two sample-level metrics for a more fine-grained assessment of GIA reconstruction outcomes. To ensure rigorous auditing, we combine two similarity metrics to define an indicator function for identifying whether a single reconstructed sample compromises privacy: $\mathbbm{1}\left[\textit{PSNR}(x, x') > \tau_1 \land \textit{LPIPS}(x, x') < \tau_2\right]$. Based on this, we propose the ``Batch Reconstruction Rate'' ($BRR$) and the ``Worst-Sample Reconstruction Rate'' ($WSRR$). $BRR$ measures the proportion of samples in a batch that leaks privacy, while $WSRR$ quantifies the likelihood of reconstructing the most vulnerable sample. We evaluate the consistency of the averaged $LPIPS$, $BRR$, and $WSRR$ in quantifying privacy leakage in GIAs across various batch sizes in Tab.~\ref{tab:audit}. The experiments are conducted using the CIFAR-10 dataset, with $\tau_1$ and $\tau_2$ set to 18 and 0.1, respectively. $WSRR$ is calculated using 10 independent attack initializations, and all the results are averaged over 10 random seeds.
\begin{table}[t]
\vspace{-2mm}
  \centering
  \renewcommand{\arraystretch}{0.75}
\setlength\tabcolsep{0.38pt}
  \caption{Comparison of Different Metrics.}
  \footnotesize
  \setlength{\tabcolsep}{2pt} % 减小列之间的空白
  \begin{tabular}{ccccccccc}
    \toprule
    \multirow{2}[4]{*}{\textbf{GIA}} & \multirow{2}[4]{*}{\textbf{Metric}} & \multicolumn{7}{c}{\textbf{Batch size}} \\
\cmidrule{3-9}          &       & 5     & 10    & 15    & 20    & 50    & 80    & 100 \\
    \midrule
    \multirow{3}[6]{*}{O} & LPIPS & 0.0338 & 0.0653 & 0.0789 & 0.0932 & \textbf{0.1167} & \textbf{0.1217} & \textbf{0.1225} \\
\cmidrule{2-9}          & BRR   & 67\%  & 34.75\% & 15.42\% & 4.06\% & \textbf{0\%} & \textbf{0\%} & \textbf{0\%} \\
\cmidrule{2-9}          & WSRR  & 95\%  & 98.75\% & 83.75\% & 47.50\% & \textbf{0\%} & \textbf{0\%} & \textbf{0\%} \\
    \midrule
    \multirow{3}[6]{*}{L} & LPIPS & 0.0446 & 0.0706 & 0.0875 & 0.0973 & \textbf{0.1238} & \textbf{0.1295} & \textbf{0.1321} \\
\cmidrule{2-9}          & BRR   & 48.25\% & 25.63\% & 11.67\% & 5.13\% & \textbf{0.27\%} & \textbf{0.11\%} & \textbf{0.10\%} \\
\cmidrule{2-9}          & WSRR  & 82.50\% & 87.50\% & 63.75\% & 55\%  & \textbf{7.50\%} & \textbf{6.25\%} & \textbf{8.75\%} \\
    \bottomrule
    \end{tabular}%
  \label{tab:audit}%
  \vspace{-3mm}
\end{table}%
Our findings demonstrate that the averaged $LPIPS$ effectively measures privacy leakage in GIAs. When $LPIPS$ is below 0.1, both $BRR$ and $WSRR$ are high, indicating significant privacy risks. Conversely, when $LPIPS$ approaches or exceeds 0.1, $BRR$ approaches zero, and even the worst samples cannot be reliably reconstructed.

\noindent\textbf{\Rmnum{3}. Introduction to post-processings: }\textbf{(1) Quantization} refers to transforming gradients to lower precision (e.g., 4-$bit$, 1-$bit$) before sharing. We consider the quantization method QSGD ([1, 2, 3, 4]-$bit$) in \cite{alistarh2017qsgd} and SignSGD\cite{bernstein2018signsgd}. \textbf{(2) Sparsification} transforms a full gradient to a sparse one with a subset of significant elements and sets others to zero\cite{zhao2023towards}. We consider $top-k$ sparsification\cite{eghlidi2020sparse}, selecting proportion $k$ ([0.6, 0.7, 0.8, 0.9, 0.95]) greatest absolute values as significant elements. \textbf{(3) Clipping} ensures the gradient values are all in a predefined bound by clipping extreme values. Here, we consider flat clipping\cite{mcmahan2017learning} with bound ([0.5, 0.3, 0.1, 0.05, 0.01] for ResNet-18, [1, 0.7, 0.5, 0.3, 0.1] for Swin). \textbf{(4) Perturbation} refers to adding noise into the gradients. In FL, clients often locally perturb the gradients before sharing to achieve local-differential-privacy\cite{naseri2020local}. Here, we add Gaussian noise with multipliers ([0.05, 0.1, 0.15, 0.2, 0.25] for ResNet-18, [0.005, 0.007, 0.01, 0.03, 0.05] for Swin) to the gradient.\par

\noindent\textbf{\Rmnum{4}. Traditional defense mechanisms in FL systems:}  We provide a discussion on traditional defense mechanisms against GIAs in FL, focusing on their privacy-utility tradeoffs, as well as their computational and communication overhead.

\textbf{$\bullet$ Differential Privacy (DP).} DP ensures that individual data samples from local datasets are hard to identify or infer by applying perturbation mechanisms, such as adding Gaussian noise, to the shared gradients~\cite{geyer2017differentially,wei2020federated,hu2023shield}. However, achieving an optimal privacy-utility tradeoff when applying DP in FL is challenging~\cite{wang2022protect}. To effectively defend against GIAs, significant perturbations are often required, which can severely degrade FL training accuracy~\cite{hu2024does}. Additionally, introducing noise at the client slows the training process, necessitating more communication rounds for convergence~\cite{hu2023federated}.

\textbf{$\bullet$ Secure Multi-Party Computation (SMPC).} SMPC enables collaborative computations among multiple participants without disclosing private inputs to other participants, thereby ensuring privacy preservation~\cite{bonawitz2017practical,mohassel2017secureml}. In FL, SMPC-based secure aggregation protocols protect client gradients by employing techniques such as Secret Sharing and pairwise masking. For instance, local gradients are masked through the weighted average of gradient vectors from a random subset, and the random factors cancel out during aggregation by a trusted server~\cite{bonawitz2017practical}. As a result, for honest-but-curious servers without additional side information, SMPC effectively defends against GIAs by revealing only aggregated gradients rather than individual ones. Nevertheless, SMPC introduces significant computational and communication overhead. For example, in Secret Sharing, all shares generated by one client are required to interact with other clients.\, and this overhead grows exponentially with the number of clients~\cite{zhang2022security}. 

\textbf{$\bullet$ Homomorphic Encryption (HE).} HE allows specific computations, such as addition, to be performed directly on encrypted gradients without requiring decryption~\cite{zhang2020batchcrypt,cheng2021secureboost}. By performing gradient aggregation on ciphertexts, HE ensures that gradients remain inaccessible to external parties, including the server. Unlike DP, HE does not reduce training accuracy, as no obfuscation is added to the gradients during encryption. However, HE imposes substantial computational and communication overhead due to the complexity of encrypting gradients and transmitting the resulting ciphertexts.

\noindent\textbf{\Rmnum{5}. Additional results in Sec.~\ref{sec:post-processing}:}
\begin{table}[H]
%\vspace{-2mm}
  \centering
  \renewcommand{\arraystretch}{0.75}
\setlength\tabcolsep{0.38pt}
  \caption{The ROG\cite{yue2023gradient} against Post-Processing Techniques.}
  \footnotesize
  \setlength{\tabcolsep}{2pt} % 减小列之间的空白
  \begin{tabular}{cccccc}
    \toprule
    \multirow{2}[4]{*}{\textbf{Model}} & \multirow{2}[4]{*}{\textbf{Dataset}} & \multicolumn{4}{c}{\textbf{Post-processing}} \\
\cmidrule{3-6}          &       & Q     & S     & C     & P \\
    \midrule
    \multirow{2}[4]{*}{ResNet-18} & C10   & \textbf{0.1548} & \textbf{0.1823} & \textbf{0.1771} & \textbf{0.229} \\
\cmidrule{2-6}          & C100  & \textbf{0.1168} & \textbf{0.1144} & 0.0529 & \textbf{0.1645} \\
    \midrule
    \multirow{2}[4]{*}{Swin} & C10   & \textbf{0.1680} & \textbf{0.1242} & \textbf{0.1117} & \textbf{0.1900} \\
\cmidrule{2-6}          & C100  & \textbf{0.1820} & 0.0906 & 0.045 & \textbf{0.1856} \\
    
    \bottomrule
    \end{tabular}%
  \label{tab:rog}%
 \vspace{-3mm}
\end{table}%

\noindent\textbf{\Rmnum{6}. Systematization on gradient inversion attacks.}

\begin{table}[htbp]
\centering

\renewcommand{\arraystretch}{0.1}
\setlength\tabcolsep{0.5 pt}
\tiny
%\scriptsize
  \caption{Systematization on Gradient Inversion Attacks.}
    \begin{threeparttable}
    \begin{tabular}{lccclccl}
    \toprule
    \multicolumn{1}{c}{\multirow{2}[4]{*}{\textbf{Publication}}} & \multicolumn{4}{c}{\textbf{Threat mode}} & \multicolumn{2}{c}{\textbf{Attack}} & \multicolumn{1}{c}{\multirow{2}[4]{*}{\extracolsep{5pt} \textbf{Defence\tnote{4}}}} \\
\cmidrule{2-7}       & \multicolumn{1}{c}{\textbf{Server’s Trust.\tnote{1}}} & \textbf{Capability} & \textbf{Goal} & \multicolumn{1}{c}{\textbf{Assumption\tnote{2}}} & \extracolsep{3pt} \textbf{Strategy\tnote{3}} & \textbf{Modality}  & \\
    \midrule
    Wang et al. (2019)\cite{wang2019beyond} & \multicolumn{1}{c}{HBC, M} & Active, Passive & data  & 0, 1, 2 & GIA-O & CV  & None  \\
    \midrule
    Zhu et al. (2019)\cite{zhu2019deep} & HBC   & Passive & label, data & 0     & GIA-O     & CV, NLP  & \textbf{Q, S, P} \\
    \midrule
    Zhao et al. (2020)\cite{zhao2020idlg} & HBC   & Passive & label, data & 0     & GIA-O    & CV  & None \\
    \midrule
    Geiping et al. (2020)\cite{geiping2020inverting} & HBC   & Passive & data  & 0, 1, 3 & GIA-O   & CV  & None  \\
    \midrule
    Fan et al. (2020)\cite{fan2020rethinking} & HBC   & Passive & label, data & 0, 1  & GIA-A  & CV  & \textbf{P}  \\
    \midrule
    Zhu et al. (2021)\cite{zhu2020r} & HBC   & Passive & data  & 0, 1  & GIA-A     & CV & \textbf{P}   \\
    \midrule
    Wainakh et al. (2021)\cite{wainakh2021user} & HBC   & Passive & label & 0     & GIA-A    & CV   & \textbf{S, P}  \\
    \midrule
    Geng et al. (2021)\cite{geng2021towards} & HBC   & Passive & label, data & 0, 1, 3  & GIA-O   & CV & None   \\
    \midrule
    Deng et al. (2021)\cite{deng2021tag} & HBC   & Passive & data  & 0, 1  & GIA-O     & NLP & None   \\
    \midrule
    Jeon et al. (2021)\cite{jeon2021gradient} & HBC   & Passive & data  & 0, 2 & GIA-L & CV  & \textbf{S, P}  \\
    \midrule
    Yin et al. (2021)\cite{yin2021see} & HBC   & Passive & label, data & 0, 1, 4 & GIA-O     & CV   & None \\
    \midrule
    Dang et al (2021)\cite{dang2021revealing} & HBC   & Passive & label & 0     & GIA-O, GIA-A  & CV, NLP & \textbf{Q, S}  \\
    \midrule
    Li et al. (2022)\cite{li2022auditing} & HBC   & Passive & label, data & 0, 1, 2 & GIA-L & CV & \textbf{S, P, C}  \\
    \midrule
    Hatamizadeh et al. (2022)\cite{hatamizadeh2022gradvit} & HBC   & Passive & data  & 0, 1, 2, 4 & GIA-O    & CV & None   \\
    \midrule
    Lu et al. (2022)\cite{lu2022april} & HBC   & Passive & data  & 0, 1  & GIA-O, GIA-O & CV  & \textbf{P}    \\
    \midrule
    Balunovic et al. (2022)\cite{balunovic2022lamp} & HBC   & Passive & data  & 0, 1, 2 & GIA-L & NLP  & \textbf{P}  \\
    \midrule
    Gupta et al. (2022)\cite{gupta2022recovering} & HBC   & Passive & data  & 0, 1  & GIA-O  & NLP & \textbf{S, P}     \\
    \midrule
    Dimitrov et al. (2022)\cite{dimitrov2022data} & HBC   & Passive & label, data & 0, 1, 3 & GIA-O      & CV & None    \\
    \midrule
    Xu et al. (2022)\cite{xu2022agic} & HBC   & Passive & label, data & 0, 1, 3, 4 & GIA-O     & CV   & None   \\
    \midrule
    Ma et al. (2023)\cite{ma2022instance} & HBC   & Passive & label & 0     & GIA-A    & CV  & None    \\
    \midrule
    Li et al. (2023)\cite{li2023temporal} & HBC   & Passive & label, data & \multicolumn{1}{p{4.625em}}{0, 4} & GIA-O    & CV, NLP & None  \\
    \midrule
    Vero et al. (2023)\cite{vero2022data} & HBC   & Passive & label, data & 0     & GIA-O, GIA-A & Tabular & None  \\
    \midrule
    Lam et al. (2021)\cite{lam2021gradient} & M     & Active & data  & 0, 5   & GIA-A     & CV   & None   \\
    \midrule
    Boenisch et al. (2021)\cite{boenisch2023curious} & M     & Active & data  & 0,5   & GIA-A     & CV   & \textbf{C, P}   \\
    \midrule
    Fowl et al. (2021)\cite{fowl2021robbing} & M     & Active & data  & 0, 5   & GIA-A     & CV  & None   \\
    \midrule
    Chu et al. (2022)\cite{chu2022panning} & M     & Active & data  & 0, 5   & GIA-A     & NLP  & None   \\
    \midrule
    Fowl et al. (2022)\cite{fowl2022decepticons} & M     & Active & data  & 0, 5   & GIA-A     & NLP  & \textbf{C, P}   \\
    \midrule
    Pasquini et al. (2022)\cite{pasquini2022eluding} & M     & Active & label, data & 0, 5   & GIA-O     & CV & None  \\
    \midrule
    Wen et al. (2022)\cite{wen2022fishing} & M     & Active & data  & 0, 5   & GIA-A     & CV  & \textbf{Q}  \\
    \midrule
    Zhao et al. (2023)\cite{zhao2023resource} & M     & Active & data  & 0,5   & GIA-A     & CV    & None \\
    \midrule
    Fang et al. (2023)\cite{fang2023gifd} & HBC     & Passive & data  & 0,1,2   & GIA-L     & CV    & \textbf{S, C, P} \\
    \midrule
    Yue et al. (2023)\cite{yue2023gradient} & HBC     & Passive & data  & 0,1,2   & GIA-L     & CV    & \textbf{Q, S, P} \\
    \midrule
    Garov et al. (2024)\cite{garov2023hiding} & M     & Active & data, label  & 0,2   & GIA-L     & CV    & \textbf{C, P} \\
    \midrule
    Xiong et al. (2024)\cite{xiong2024gi} & HBC     & Passive & data  & 0,1,2   & GIA-L     & CV    & \textbf None \\
    \midrule
    Liu et al. (2024)\cite{liu2024gradient} & HBC     & Passive & data,label  & 0   & GIA-O     & CV    & \textbf{P} \\
    \midrule
    Wang et al. (2024)\cite{wang2024breaking} & M     & Active & label  & 0,5   & GIA-A     & CV    & \textbf{S, P} \\
    \midrule
    Dimitrov et al. (2024)\cite{dimitrov2024spear} & HBC     & Passive & data  & 0   & GIA-A     & CV    & None \\
    \midrule
    Wang et al. (2024)\cite{wang2024towards} & HBC     & Passive & label  & 0   & GIA-A     & CV    & \textbf{P} \\
    \bottomrule
    \end{tabular}%
    \begin{tablenotes}    % 添加命令mini-batch label
        \tiny
        %\footnotesize               % 添加命令
        \item[1] HBC for Honest-but-curious; M for Malicious
        % \item[2] For Label reconstruction: Class-wise: infer how many classes of labels are inside a batch\cite{yin2021see,dang2021revealing,dimitrov2022data}. Instance-wise: infer how many instances are in each class\cite{ma2022instance}. 
        \item[2] Assumption: [0] Basic information, [1] Priors, [2] Data distribution, [3] Client-side training details, [4] BN statistics, [5] Malicious behavior
        \item[3] GIA-O for GIA with Observable Space Optimization; GIA-L for GIA with Latent Space Optimization; GIA-A for Analytic-based GIA	
        \item[4] Quantization \textbf{(Q)}, Sparsification \textbf{(S)}, Clipping \textbf{(C)}, and Perturbation \textbf{(P)} 
      \end{tablenotes}   
    \end{threeparttable}
  \label{tab:sys}%
\end{table}%

\end{document}